# KOOPMAN-Luenberger Observer Design for Nonlinear Systems with Application to the Monitoring of a Latent Thermal Energy Storage

Mustapha Habib[1], Dario Aguiar[2], Esther Kieseritzky[2], Tilman Barz[3] and Qian Wang[1]

[1] Division of Building Technology and Design, Department of Civil and Architectural Engineering, KTH Royal Institute of Technology, 11428 Stockholm, Sweden

[2] Rubitherm Technologies GmbH, Sperenberger Str. 5A, D-12277 Berlin

[3] AIT Austrian Institute of Technology GmbH, Center for Energy, Giefinggasse 2, 1210 Vienna, Austria

Corresponding author: Mustapha habib (mushab@kth.se)

**Abstract** - State estimation for nonlinear dynamical systems remains a fundamental challenge, particularly when measurements are sparse and internal states are inaccessible. This work presents a KOOPMAN-based Linear State Observer (KOOPMAN-LSO) design framework that enables linear observer synthesis for nonlinear systems through KOOPMAN operator theory. The nonlinear dynamics are lifted into a higher-dimensional observable space using physics-informed basis functions, where a linear predictor with control is identified via extended dynamic mode decomposition with control (eDMDc). A discrete-time Luenberger observer is then constructed in the lifted space, and the observer gain is obtained through a dual linear - quadratic regulator (LQR) formulation to ensure stable and tunable estimation error dynamics.

The proposed framework combines the representational capability of KOOPMAN lifting with the simplicity and computational efficiency of linear observer design, providing a systematic approach for nonlinear state estimation under limited sensing. Its effectiveness is demonstrated on a latent thermal energy storage (LTES) system based on phase-change materials (PCM), where internal temperature states are not directly measurable. Experimental results under varying operating conditions show accurate reconstruction of unmeasured states from limited output measurements, illustrating the potential of KOOPMAN-LSO design for practical nonlinear systems. The proposed approach achieves high-fidelity reconstruction with an RMSE as low as 0.0819 °C for the LTES outlet temperature and generally below 1.0 °C for observable internal PCM temperatures.

## 1. Introduction

### 1.1. Background

Data-driven modeling approaches have emerged as a promising alternative to first-principles models for complex thermal systems, particularly when accurate physical parameterization is difficult or computationally prohibitive. Among these methods, Dynamic Mode Decomposition (DMD) and its input-aware extension, DMD with control (DMDc), provide linear approximations of nonlinear dynamics directly from data. Extended DMD (eDMD) further generalizes this framework by lifting the system into a higher-dimensional space of nonlinear observables, yielding a finite-dimensional approximation of the KOOPMAN operator. This KOOPMAN-based representation enables the recasting of nonlinear dynamics into a linear form in lifted coordinates, thereby facilitating analysis and control design while preserving essential nonlinear behavior. In the context of latent thermal energy storage, such approaches allow the incorporation of physics-informed observables related to phase transition phenomena, offering a structured yet flexible modeling framework that bridges purely empirical models and full-order physics-based descriptions.

While KOOPMAN-based models provide a linear structure for nonlinear systems, their practical utility in monitoring applications depends critically on the ability to reconstruct unmeasured states from available outputs. This motivates the design of state observers operating in the lifted coordinate space, where linear observer theory can be exploited despite the underlying nonlinear dynamics. However, observability in KOOPMAN coordinates is nontrivial, as the lifted state dimension may exceed the number of physical states and measurements, and only a subset of observables may be directly related to measured outputs. Luenberger-type observers, particularly when combined with optimal gain design techniques such as linear quadratic regulation (LQR), offer a systematic approach to ensure stable estimation error dynamics under these conditions. By embedding physical measurements into the lifted space and enforcing observability through appropriate output mappings, KOOPMAN-based observers enable progressive convergence of estimated latent thermal states, providing a viable solution for real-time monitoring of strongly nonlinear thermal energy storage systems.

Latent thermal energy storage (LTES) systems based on phase change materials (PCMs) are increasingly adopted to enhance energy efficiency and operational flexibility in thermal management applications. Despite their advantages, accurate monitoring of internal thermal states remains a major challenge due to the strongly nonlinear heat transfer mechanisms induced by phase change, hysteresis effects, and spatially distributed temperature gradients. In practical installations, only a limited number of measurements - typically inlet or outlet fluid temperatures - are available, while the internal PCM temperatures that govern the stored and released energy remain largely unmeasured. This limited sensing capability, combined with nonlinear and time-varying dynamics, complicates state reconstruction and hampers the deployment of advanced control strategies. Consequently, there is a clear need for efficient, reliable, and scalable monitoring solutions capable of estimating latent thermal states using sparse measurements under realistic operating conditions.

### 1.1. Previous works

As stated earlier, accurate monitoring of latent thermal energy storage (LTES) systems based on PCMs is inherently challenging due to the strongly nonlinear and distributed nature of the phase transition process. The coexistence of solid and liquid phases, temperature plateaus during melting and solidification, and hysteresis effects result in dynamics that are difficult to capture using sparse boundary measurements alone. In practical systems, direct access to internal PCM temperatures or phase distributions is limited, motivating extensive research into indirect monitoring strategies that estimate internal thermal states or the state of charge (SoC) from a small number of sensors. Early work primarily relied on dense temperature sensor networks embedded within the PCM domain to reconstruct thermal fields and infer melting fronts [1–3]. Still, such approaches are often impractical in large-scale or commercial systems due to cost, reliability, and intrusiveness.

To overcome these limitations, physics-based modeling approaches have been widely investigated for state reconstruction in PCM-based TES. High-fidelity numerical models, including finite-volume and computational fluid dynamics (CFD) formulations with enthalpy methods, have been used to simulate internal temperature fields and phase boundaries with high accuracy [4–6]. While these models provide valuable insight and serve as virtual sensors, their computational complexity typically precludes real-time monitoring applications. Consequently, reduced-order and lumped-parameter models have been proposed, enabling observer-based estimation of internal states using limited measurements such as inlet and outlet fluid temperatures [7–9]. In this context, nonlinear observers and Kalman filter variants have been developed to estimate both temperature distributions and latent energy content, demonstrating convergence under realistic operating conditions [10–12]. These approaches explicitly exploit physical conservation laws but often require careful parameter identification and may suffer from performance degradation when unmodeled dynamics or parameter drift are present.

More recently, data-driven and hybrid estimation techniques have gained increasing attention as a means of addressing modeling uncertainty and structural nonlinearities in PCM systems. Statistical learning methods, including regression-based SoC estimation and Gaussian process models, have been employed to infer latent thermal states directly from historical operational data [13–15]. Machine learning approaches such as artificial neural networks and recurrent architectures have also been explored for LTES monitoring, particularly when large datasets are available [16,17]. However, purely data-driven models often lack interpretability and robustness outside the training domain. This has motivated the development of hybrid strategies that combine reduced-order physical models with data-driven corrections or observers, enabling improved generalization while maintaining physical consistency [18–20].

Within this broader landscape, KOOPMAN-based and DMD-type methods represent a unifying framework that bridges physics-based and data-driven estimation by lifting nonlinear PCM dynamics into a linear representation suitable for observer design, offering a promising direction for scalable and real-time monitoring of latent thermal energy storage systems. Introducing the KOOPMAN operator as a linear operator acting on observable functions of Hamiltonian systems establishes the theoretical basis for linear representations of nonlinear dynamics, though it lacked a computational framework. The modern revival by [22] applied KOOPMAN spectral analysis to fluid dynamics, demonstrating practical applications but facing computational complexity. A crucial connection was made by [23], linking Dynamic Mode Decomposition (DMD) to KOOPMAN theory, providing a data-driven approximation framework, albeit initially limited to autonomous systems.

The development of Extended Dynamic Mode Decomposition (eDMD) marked a significant advance. [24] extended DMD to include nonlinear observables via dictionary functions, enabling finite-dimensional approximations but grappling with the curse of dimensionality. For control applications, [25] formulated eDMD with control (eDMDc), creating linear predictors for nonlinear systems in lifted spaces, though it required persistent excitation for accurate identification. Concurrently, [26] developed sparse regression techniques (SINDy) to identify parsimonious KOOPMAN-invariant subspaces, offering automated model discovery but showing sensitivity to measurement noise.

For state estimation, several KOOPMAN-based methods emerged. [27] combined KOOPMAN linearization with Kalman filtering, creating the KOOPMAN Kalman Filter for estimation in lifted coordinates, which assumed Gaussian noise distributions. [28] explored local KOOPMAN operators for data-driven control of robotic systems, providing stability guarantees but requiring multiple local models. A paradigm shift occurred with the integration of deep learning, as shown by [29], which used autoencoders to learn KOOPMAN eigenfunctions automatically, overcoming manual basis selection at the cost of interpretability and large data requirements.

Specific applications to thermal energy storage reveal domain-specific challenges and solutions. [30] employed neural operators to learn solution operators for parametric heat transfer PDEs, achieving efficiency but requiring extensive training data. For PCM systems, [31] used Proper Orthogonal Decomposition for model order reduction, achieving high accuracy with reduced models, though results were limited to specific geometries. The complex dynamics of phase change were addressed by [32] using adaptive moving boundary methods for precise phase front tracking, which incurred high computational costs for 3D problems.

Observer design for thermal systems has been approached from various angles. [33] designed high-gain observers for nonlinear heat conduction, achieving robustness to model uncertainties but requiring precise knowledge of the nonlinearity structure. [34] reviewed Moving Horizon Estimation, an optimization-based method effective for handling constraints and noise but computationally demanding. For tunable and stable estimation, [35] utilized LQR-based observer design, providing guaranteed convergence but dependent on accurate linear models.

Hybrid physics-data approaches represent a promising frontier. [36] pioneered Physics-Informed Neural Networks (PINNs), constraining neural networks with physical laws for data-efficient learning, though training could be unstable. [37] developed projection-based model reduction formulations for physics-based machine learning, preserving interpretability but limited to systems with known equation structures. To automate basis selection, [38] introduced adaptive dictionary learning for eDMD, improving generalization at the cost of computational overhead for dictionary optimization.

Comparative and validation studies provide critical benchmarks. [39] conducted a systematic comparison of data-driven methods for convective heat transfer, finding KOOPMAN-based methods offered the best accuracy-computation trade-off, though the study was simulation-only. [40] provided crucial experimental validation of data-driven thermal models on laboratory-scale systems, demonstrating practical applicability but highlighting limited scalability

### 1.2. Proposed work

In this work, a KOOPMAN-based Linear State Observer (KOOPMAN-LSO) framework is developed for the estimation of internal temperature states in nonlinear dynamical systems, with a specific focus on LTES systems based on PCMs. The proposed approach directly addresses the challenge of reconstructing distributed nonlinear thermal states from sparse boundary measurements, which remains a key limitation in existing monitoring strategies.

The nonlinear system is lifted into a higher-dimensional observable space using physics-informed basis functions, enabling a finite-dimensional approximation of the KOOPMAN operator. In contrast to purely data-driven lifting approaches such as deep autoencoder-based KOOPMAN models [29], which require large datasets and lack interpretability, the proposed formulation embeds prior physical knowledge of heat transfer and phase transition processes into the observable selection. This improves generalization and reduces model complexity while preserving essential nonlinear dynamics. A linear predictor with control is then identified using eDMDc [25], allowing the incorporation of system inputs and enabling applicability under varying operating conditions - an aspect not addressed in early KOOPMAN-DMD formulations [23,24].

For state estimation, a discrete-time Luenberger observer is constructed in the lifted space. Unlike KOOPMAN-based Kalman filtering approaches [27], which rely on stochastic noise assumptions and covariance tuning, the observer gain in this work is obtained through a dual Linear Quadratic Regulator (LQR) formulation. This provides a deterministic and systematically tunable design ensuring stable estimation error dynamics. Compared to high-gain observers [33] or moving horizon estimation methods [34], the proposed observer retains low computational complexity while avoiding sensitivity to noise amplification or the need for online optimization.

From an application perspective, the proposed framework departs from conventional physics-based estimation methods for PCM systems [7–12], which rely on reduced-order models and often suffer from parameter sensitivity and limited adaptability. It also addresses the limitations of purely data-driven approaches [13–17], which may lack robustness outside the training domain. In contrast, the KOOPMAN-LSO framework offers a hybrid structure, combining data-driven identification with physically meaningful representations, thereby ensuring both interpretability and predictive capability. Furthermore, compared to recent hybrid and learning-based methods such as PINNs [36] or neural operators [30], the proposed approach achieves significantly lower computational overhead, making it suitable for real-time monitoring.

The effectiveness of the framework is demonstrated experimentally on a PCM-based LTES system under varying charging and discharging conditions. The main contributions of this work can be summarized as follows:

- The development of a KOOPMAN-based observer design framework with control inputs, enabling linear estimation of nonlinear thermal systems.
- The introduction of physics-informed lifting for PCM dynamics, improving model interpretability and robustness compared to the purely data-driven KOOPMAN approach.
- The formulation of a Luenberger observer with LQR-based gain design in the lifted space, providing stable and tunable estimation without stochastic assumptions.
- A comprehensive experimental validation demonstrates accurate and real-time reconstruction of internal PCM temperature states under realistic operating conditions.

## 2. Method

In this section, we explore the mathematical foundation of the proposed KOOPMAN-based Luenberger observer, starting from the DMDc and its extended version (eDMDc), ending up with the linear Luenberger observer built on the KOOPMAN linear representation on a finite lifted observation state.

### 2.1. DMD problem formulation

Consider a discrete-time nonlinear system with control input:

$$\begin{aligned} x_{k+1} &= f(x_k, u_k) \\ y_k &= h(x_k) \end{aligned} \tag{1}$$

Where:

- $x_k \in \mathbb{R}^{n_x}$ is the physical state.
- $u_k \in \mathbb{R}^{n_u}$ is the input.
- $y_k \in \mathbb{R}^{n_y}$ is the measured output.

The system is nonlinear, partially observed, and not directly suitable for linear observer design. Therefore, we aim, in this study, to find a linear approximation using a data-driven strategy.

DMD has been extensively used as an efficient data-driven approach in various applications [6]. Its main powerful feature is the ability to find approximate linear representations of unknown system dynamics based on measurement data. DMD is based on collecting pairs of snapshots of system states as they evolve in time. These snapshots include system internal states, which are arranged into matrices $X$ and $X'$:

$$X = \begin{bmatrix} | & | & \dots & | \\ x(t_1) & x(t_2) & \dots & x(t_{m-1}) \\ | & | & \dots & | \end{bmatrix} \quad X' = \begin{bmatrix} | & | & \dots & | \\ x(t_2) & x(t_3) & \dots & x(t_m) \\ | & | & \dots & | \end{bmatrix} \tag{2}$$

The DMD algorithm seeks the leading spectral decomposition of the best-fit linear operator $A$ that relates the two snapshot matrices in time:

$$X' \approx AX \tag{3}$$

The best-fit operator $A$ then establishes a linear dynamic system that best advances snapshot measurements forward in time. If we assume uniform sampling in time, this becomes:

$$x_{k+1} = Ax_k \tag{4}$$

$A$ can typically be identified using linear regression, where the correct expression is written below. Here, † denotes the pseudo-inverse.

$$A = \arg\min_{A} \|X' - AX\| = X'X^{\dagger} \tag{5}$$

In some applications, like the one we will explore in this study, the system is overdetermined; in such a way, the system is not high-dimensional, while it was possible to record many data snapshots (see Section 2.4). This results in a very suitable situation for developing a DMD framework since the number of equations is higher than the number of variables.

Like most of the dynamic models for thermal energy systems, these models can be presented using Eq. (6), which means they can be stabilized using control inputs. In this case, additional data snapshots are needed for constructing the extended DMD model, which becomes now DMDc.

$$x_{k+1} \approx Ax_k + Bu_k \tag{6}$$

Let us define the snapshot matrix $X = [x_1 \; x_2 \; ... \; x_{m-1}]$ and the time-shifted snapshot matrix $X' = [x_2 \; x_3 \; ... \; x_m]$, a matrix for the actuation input history and disturbances are also assembled:

$$U = \begin{bmatrix} | & | & \cdots & | \\ u_1 & u_2 & \cdots & u_m \\ | & | & \cdots & | \end{bmatrix} \tag{7}$$

The dynamics in (6) may be written in terms of data matrices as:

$$X' \approx AX + BU \tag{8}$$

Since $B$ is unknown, both $A$ and $B$ must be simultaneously identified. In this case, the dynamics in (8) may be recast as:

$$X' \approx [A \;\; B] \begin{bmatrix} X \\ U \end{bmatrix} = G.\Omega \tag{9}$$

Similarly, the matrix $G = [A \quad B]$ is obtained via a linear regression as follows:

$$G \approx X'\Omega^{\dagger} \tag{10}$$

As in the normal DMD, $\Omega = [X^* \quad U^*]^*$ in our case study, is a low-dimensional data matrix, which makes it an overdetermined system of equations.

### 2.2. eDMDc lifting and KOOPMAN approximation

As stated previously, DMDc can be extended by incorporating nonlinear terms in the observation space, so it becomes eDMDc. This technique unlocks the system's nonlinearity while maintaining the overall linear representation, which is important for enabling linear control and observation theories.

One important feature of eDMDc is that it gives a finite approximation to the infinite KOOPMAN observation space, which means that, as underlined below:

$$\varphi : \mathbb{R}^{n_z} \to \mathbb{R}^{n_z},\ z_k = \varphi(x_k)$$

The KOOPMAN operator with control is then approximated via eDMDc as:

$$z_{k+1} = Az_k + Bu_k \tag{11}$$

Where:

- $A \in \mathbb{R}^{n_z \times n_z}$,
- $B \in \mathbb{R}^{n_z \times n_u}$.

This approximation is obtained from data by solving:

$$\min_{A,B} \left\| Z^+ - AZ - BU \right\|_F \tag{12}$$

With snapshot matrices: $Z = \varphi(X),\ Z^+ = \varphi(X^+)$.

Although dynamics evolve in lifted space, measurements depend only on the physical states, which can be mapped out using: $x_k = C_x z_k$, while the overall output is $y_k = Cz_k$.

Where:

- $C_x \in \mathbb{R}^{n_x \times n_z}$,
- $C \in \mathbb{R}^{n_y \times n_z}$.

### 2.3. Luenberger Observer for KOOPMAN

Since we have a linear representation on the lifted observation space for the data-driven nonlinear system, it is straightforward now to build a linear observer (LO) as Eq. (13) shows:

$$\hat{z}_{k+1} = A\hat{z}_k + Bu_k + L(y_k - C\hat{z}_k) \tag{13}$$

where:

- $\hat{z}_k$ is the lifted state estimate,
- $L \in \mathbb{R}^{n_z \times n_y}$ is the observer gain.

The estimation error on the lifted space is then formulated below:

$$e_k = z_k - \hat{z}_k \tag{14}$$

By subtracting observer dynamics from system dynamics, we get the estimation error dynamics as below:

$$\begin{aligned} e_{k+1} &= z_{k+1} - \hat{z}_{k+1} \\ &= (A\hat{z}_k + Bu_k) - (A\hat{z}_k + Bu_k + L(y_k - C\hat{z}_k)) \\ &= (A - LC)e_k \end{aligned} \tag{15}$$

Once we have found $L$ that stabilizes the estimation error dynamic, not all estimated observations are meaningful. Only the physical state estimates that exist in the original DMDc are needed, which should be recovered as:

$$\hat{x}_k = C_x \hat{z}_k \tag{16}$$

Where $C_x$ is the matrix for mapping the real physical states from the overall lifted state vector.

In real-world applications, the aptitude of a limited number of measurements to reconstruct the lifted observation estimates depends highly on the observability analysis using KOOPMAN operator $A$ and the lifted output matrix $C$. If $rank(o) = n_z$, then the lifted system is observable.

$$o = \begin{bmatrix} C \\ CA \\ CA^2 \\ \vdots \\ CA^{nz-1} \end{bmatrix} \tag{17}$$

To stabilize Eq. (16), many techniques can be employed, including the classical pole placement approach. In this study, we adopted the optimal pole placement via Linear Quadratic Regulator (LQR), where the goal is to minimize the objective function formulated in Eq. (18), addressing the estimate error conversion and the measurement prediction.

$$J = \sum_{k=0}^{\infty} e_k^T Q_e e_k + (y_k - \hat{y}_k) \tag{18}$$

This leads to solving the dual discrete Riccati equation, which determines the observation gains deterministically as follows:

$$L = PC^T (\mathrm{R_e} + CPC^T)^{-1} \tag{19}$$

- Large $Q_e$: fast error convergence, noise sensitivity
- Large $R_e$: slower convergence, smoother estimates

### 2.4. eDMDc model for PCM energy storage

In this subsection, the linear eDMDc representation for PCM-based latent thermal energy storage (LTES) is presented. The state vector incorporates all PCM temperature measurements $T_{p,i}$ in different coordinates in LTES and the outlet water temperature $T_{w,out}$ [°C]; $x = [T_{p,1}, T_{p,2}, \ldots, T_{p,n}, T_{w,out}]$ where $n$ is the total number of sensors; the chosen system inputs are the LTES inlet water temperature $T_{w,in}$ [°C] and flow rate $\dot{m}$ [l/min], $u = \begin{bmatrix} T_{w,in} & \dot{m} \end{bmatrix}$; one measurement $y$ is considered in this study, which is the last state in the state vector $T_{w,out}$ where $y = [0,\ldots,0,1]x$.

By applying the DMDc framework explained in Section 2.1, datasets recorded by these sensors were used to construct a linear representation of the PCM-TES, while still capturing the nonlinear patterns of temperature changes caused by the phase change. As stated earlier, to ensure better nonlinearity handling, the system can be lifted by including nonlinear functions (observations) to the original state vector. To effectively lift DMDc to a eDMDc, one should consider nonlinear relationships that reflect the PCM physics; therefore, two observable functions are considered in this context: the Sigmoid function to model the PCM phase change pattern from liquid to solid states and vice versa (see Eq. (20)), and the Gaussian function to model the PCM heat specific capacity change around the melting point (see Eq. (21)). These two functions correspond to each PCM temperature state $i$:

$$Sigm,i = \frac{1}{1 + e^{-\alpha(T_{p,i} - T_m)}} \tag{20}$$

$$Gaus,i = e^{-0.5(\frac{T_{p,i} - T_m}{\partial})^2} \tag{21}$$

Where $\alpha$ is the phase transition sharpness parameter, it controls how abruptly the phase transition occurs with temperature, and determines the width of the temperature interval over which the PCM transitions from solid-like to liquid-like behavior, and $\partial$ specific heat spread parameter, It controls the temperature bandwidth over which the apparent heat capacity is elevated and defines how localized the peak in effective specific heat is around $T_m$, the melting temperature.

In the end, our KOOPMAN model can be presented as follows:

$$z = \begin{bmatrix} T_{p,1} \\ Sigm,1 \\ Gaus,1 \\ T_{p,2} \\ Sigm,2 \\ Gaus,2 \\ \vdots \\ T_{p,n} \\ Sigm,n \\ Gaus,n \\ T_{w,out} \end{bmatrix} \tag{22}$$

### 2.5. KOOPMAN Linear Observer for PCM Temperature Estimation

Taking advantage of the obtained KOOPMAN-based linear representation, the state estimation problem can be naturally reformulated within a linear observer framework as explained in Section 2.3. In particular, this representation enables the systematic design of a linear Luenberger observer in the lifted state space. Figure 1 presents the state-space representation and associated observer structure employed for reconstructing the PCM temperature states from measured variables

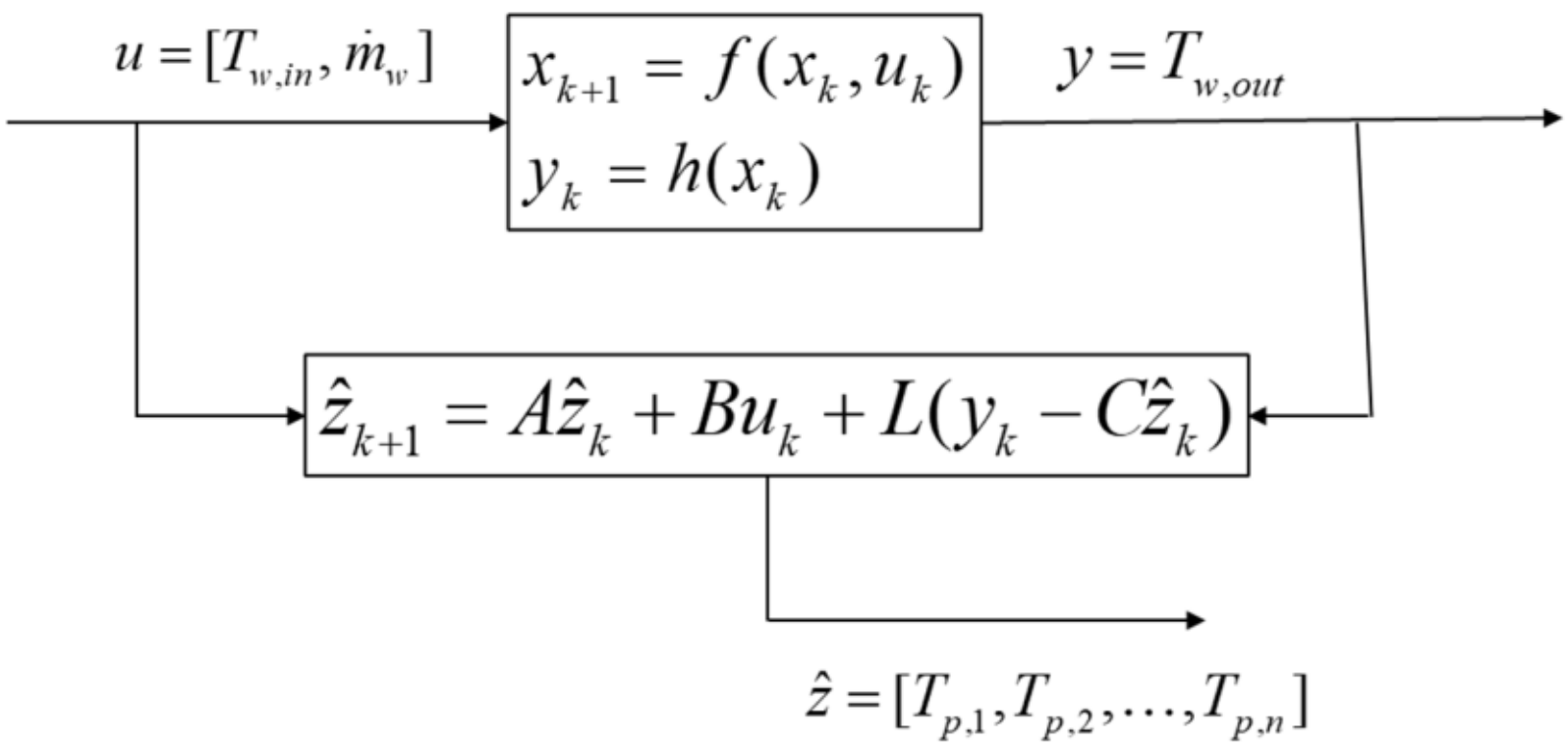

Figure 1: State-space observer framework for PCM temperature estimation in thermal energy storage

In contrast to nonlinear observer formulations - where stability and convergence often rely on local linearization, heuristic tuning, or Lyapunov-based arguments - the linear KOOPMAN model allows the observer correction gains to be computed deterministically using well-established linear control techniques, such as pole placement or LQR design.

The corrected temperature estimates can be written explicitly as:

$$T^{i}_{p,k+1} = C_T(A\hat{z}_k + Bu_k + L(y_k - C_K\hat{z}_k)) \tag{23}$$

Where:

- $\hat{z}_k$ is the estimated lifted KOOPMAN state (PCM temperatures + nonlinear observables).
- $T^i_{p,k+1}$ is the estimated PCM temperature vector.
- $A$ and $B$ are the KOOPMAN system and input matrices.
- $C_K$ is the output matrix mapping lifted states to measured outputs.
- $C_T$ is the selector matrix extracting PCM temperatures from the listed state.
- $u_K$ is the input vector.
- $L$ is Luenberger observer gain.
- $y_k - C_K\hat{z}_k$ is the estimation error.

### 2.6. Validation system description

The KOOPMAN-based observation theory presented in the study is examined in a real-world case in which it was employed to estimate and monitor the internal PCM temperatures of a PCM heat exchanger (HEX) serving as a latent TES. The implementation of the proposed methodology is found in the following two main stages:

- Stage 01: at the laboratory scale, the LTES designer and tester have the right conditions to install temperature sensors in all the LTES coordinates, which enables the generation of measurement datasets. These datasets, which include internal PCM temperatures and the hot-temperature fluid (HTF) (water in this case), enable the development of an eDMDc prediction model.
- Stage 2: at the deployment application (heating or domestic hot water), such an environment usually does not offer the right conditions to keep the same sensors utilized in Stage 1 to maintain the same monitoring performance. With this operational situation, there is a need to keep tracking the LTES performance and State-of-Charge (SoC), with a minimum set of sensors, which represents a suitable application of state observer utilization.

After multiple tests, the properties of the PCM HEX were set as follows (see Figure 2): the HEX in a module contains a capillary tube mat folded 10 times, with a total length of around 4 m per tube, as determined empirically by the manufacturer. This ensures an average pipe spacing of less than 20 mm and optimum heat exchange between HTF and PCM. For a tube length of around 4 m, the temperature of HTF at the outlet of the storage is already close to the phase change temperature, which means that there is no potential for further heat transfer. Therefore, 4 m presents the maximum usable capillary length. In contrast, for tube lengths shorter than 4 m there is still potential for energy to transfer. The 10 mats are connected in parallel with the supply and return pipe running anti-parallel. This ensures that every capillary tube in the system has the same total tube length in the system (supply pipe plus capillary tube length plus return pipe). As this connection system, known as Tichelmann or reverse return system in heating systems, is applied here, all tubes have a similar resistance with respect to the heat transfer fluid, leading to uniform exchange of heat between the transfer fluid and the PCM. Temperature differences within the storage remain as small as possible, and the melting and solidification progress are similar for each tube.

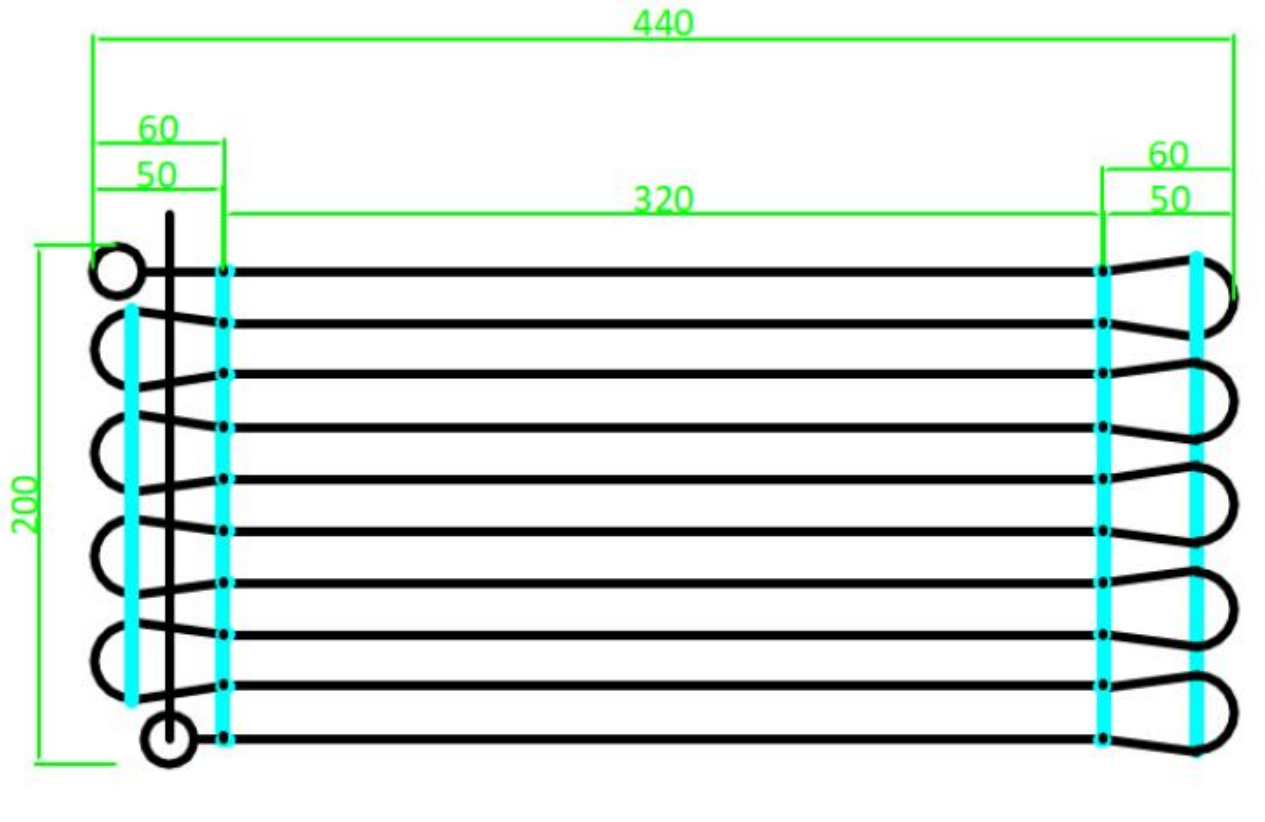

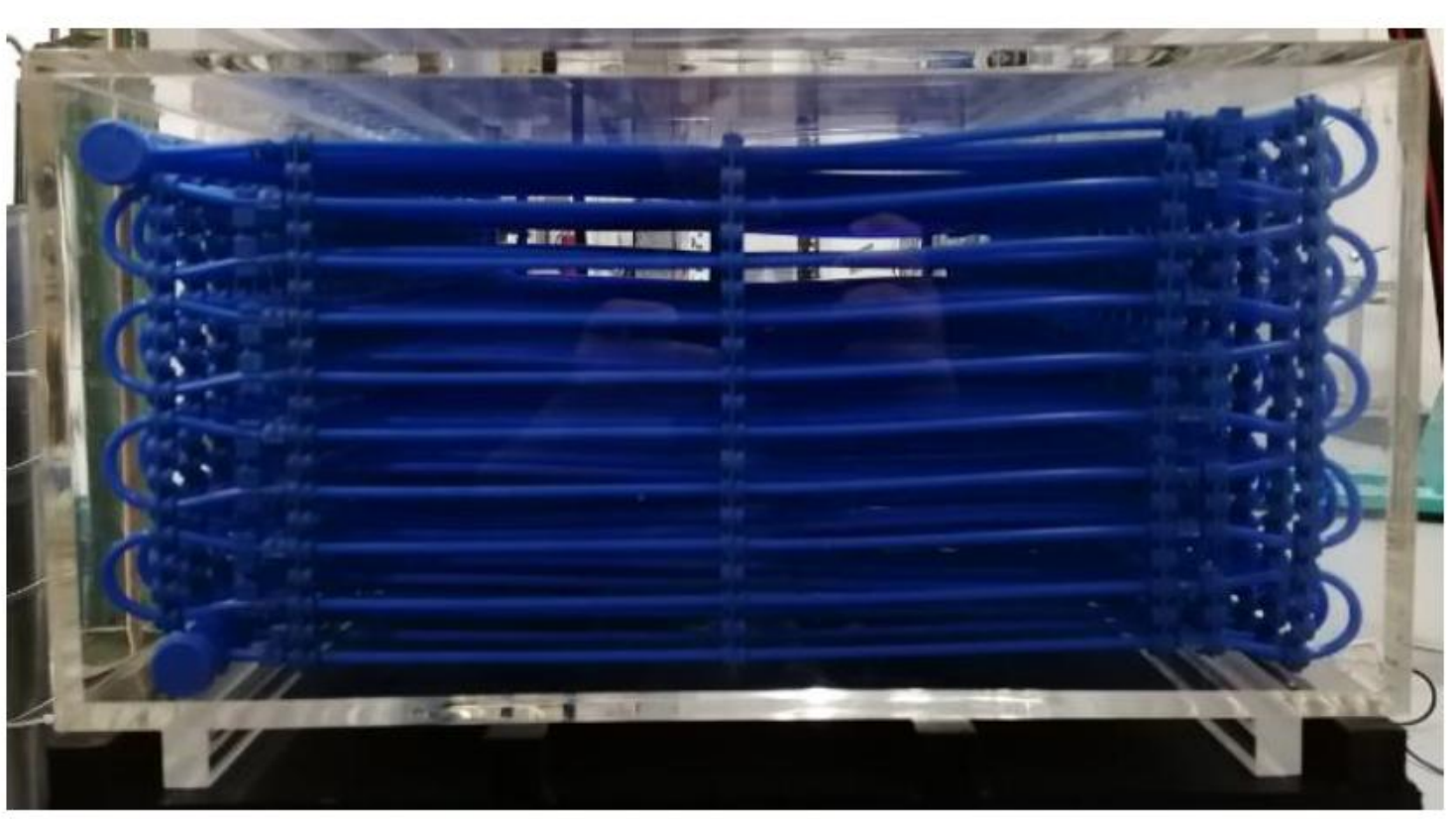

Figure 2: Design of the low-cost polymeric HEX, on the left scheme giving dimensions (in mm), on the right the realization

### 2.7. PCM material characteristics:

SP31 is an inorganic PCM belonging to the SP product series manufactured by Rubitherm Technologies GmbH. It is a salt hydrate-based material consisting of a mixture of salt, water, and additives. It has a phase transition enthalpy of 195.4 J/g and a phase change peak temperature of 32 °C and 29.5 °C for heating and cooling, respectively. It is produced and delivered as bulk material or as macroencapsulated solution in the form of Thermal Packs and Compact Storage Modules (CSM). The macroencapsulated PCM quantity in the Thermal Packs is 320 - 2400 g, while it is between 1 - 2 kg in the Compact Storage Modules. It is usually employed for active and passive cooling and heating systems, e.g., for air-conditioning, in-roof and wall elements [41].

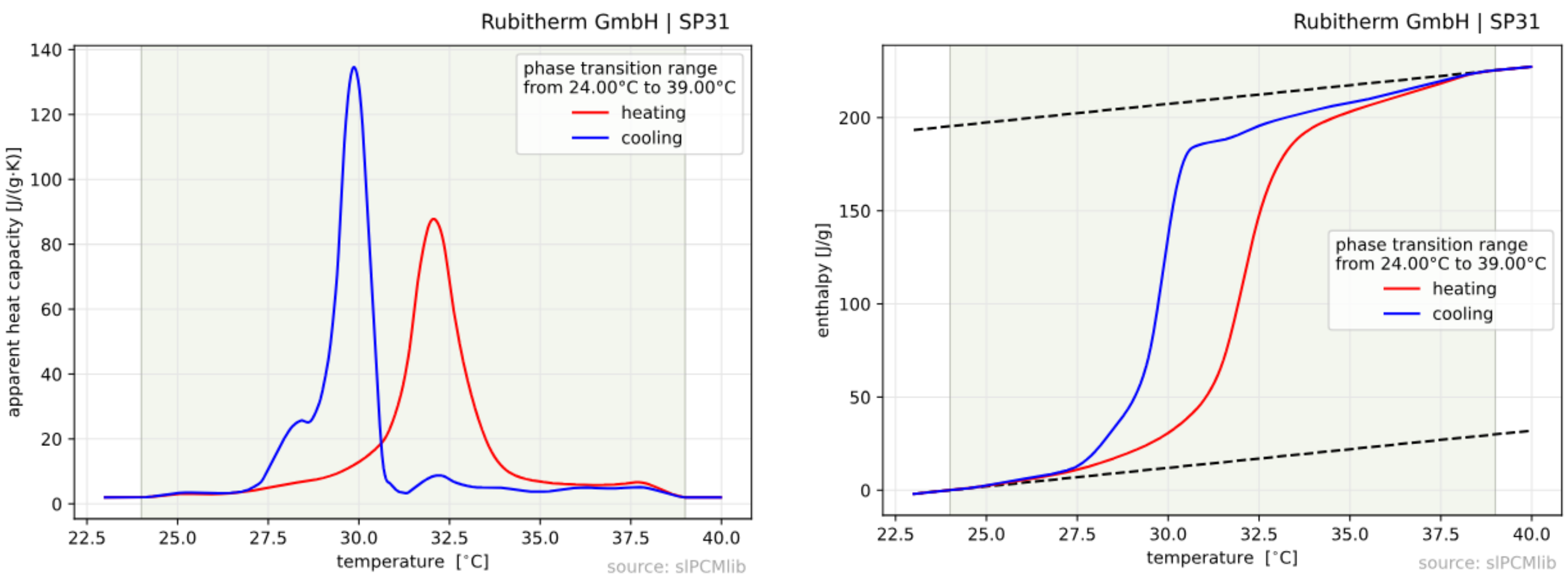

Figure 3: thermodynamic characteristics of SP31 PCM material, (left) apparent heat capacity, (right) enthalpy [41]

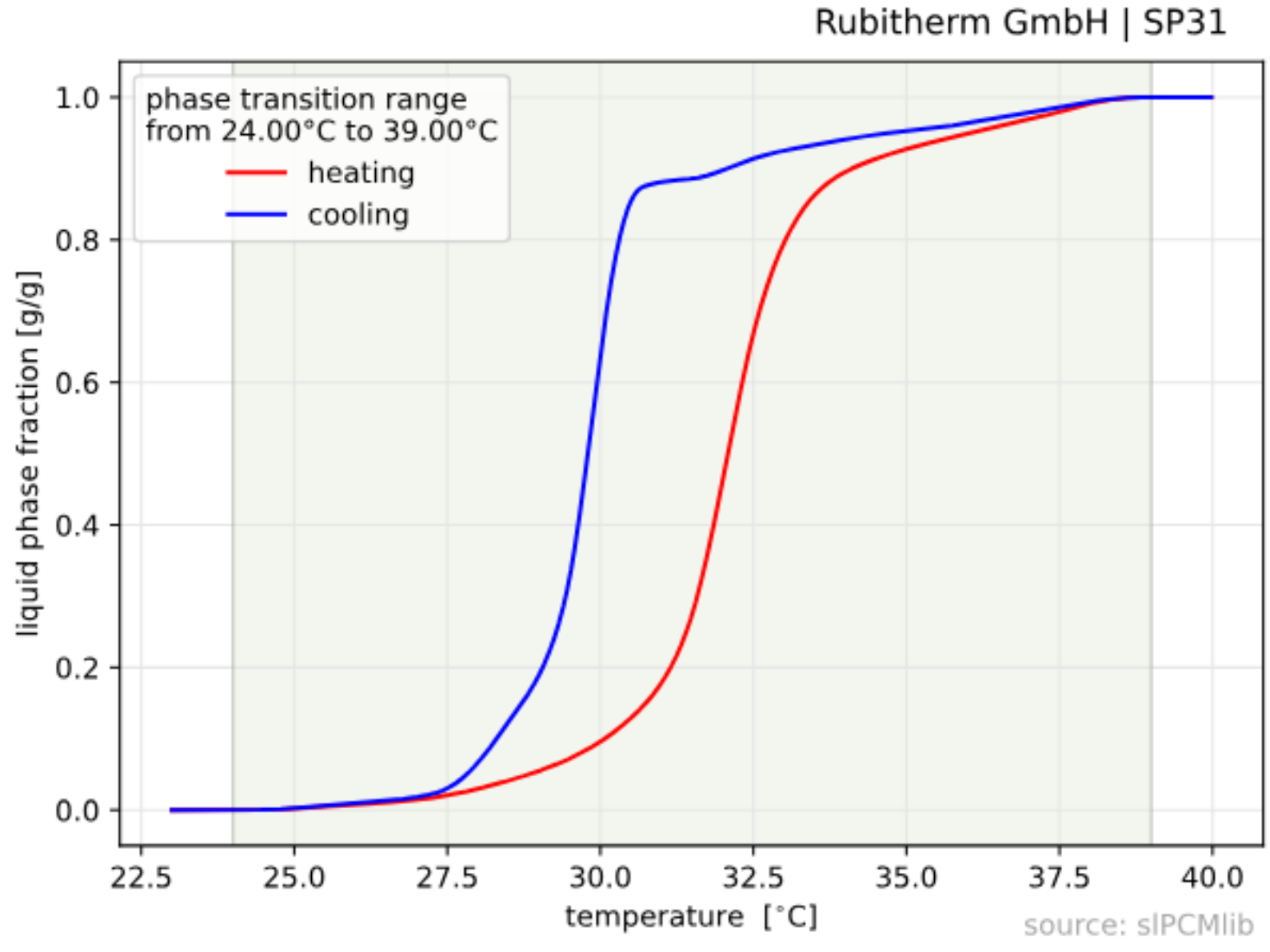

Figure 4: liquid phase fraction for SP31 PCM material for heating and cooling [41]

From Figures 3 and 4, we observe a strong nonlinearity in the PCM material characteristics, making it challenging for any linear modeling approach to accurately capture the system dynamics. KOOPMAN linear approximation is expected to tackle this thanks to the nonlinear observations that were inspired by PCM material behavior.

### 2.8. Validation procedure

Several charging and discharging cycles using a 200 l storage tank with the same HEX geometry described in Figure 2 were carried out. This allows us to evaluate the thermal properties of the module and the temperature estimates methodology developed in this study. During the tests, the discharge inlet temperature was set around 25°C for SP31 PCM material, while the charging temperature was set around 35°C. Different HTF flow rates were tested, mainly 2, 4, and 8 l/min, in order to evaluate the SoC variation pattern and the effectiveness of the KOOPMAN-LSO framework under different operation conditions. A few temperature sensors have been placed in the storage system and used to provide data for the following dynamic model architecture (see Figure 5):

- **Inputs:**
    - One sensor for the inlet temperature measurement (with a tag name B2S4)
    - One sensor for the HTF flow rate (not shown in Figure 3).
- **Measurement:**
    - One sensor for the outlet temperature measurement (with the tag number B2S3).
- **States:**
    - Nine sensors for PCM temperature measurement (with the tag names B1S1, B1S2, B1S3, B1S5, B1S6, B1S7, B1S8, B2S1, B2S2).

After considering all these measurements for the KOOPMAN model development, the proposed KOOPMAN-LSO is designed to eliminate the need for all PCM temperature sensors by providing estimates using the Luenberger observer instead.

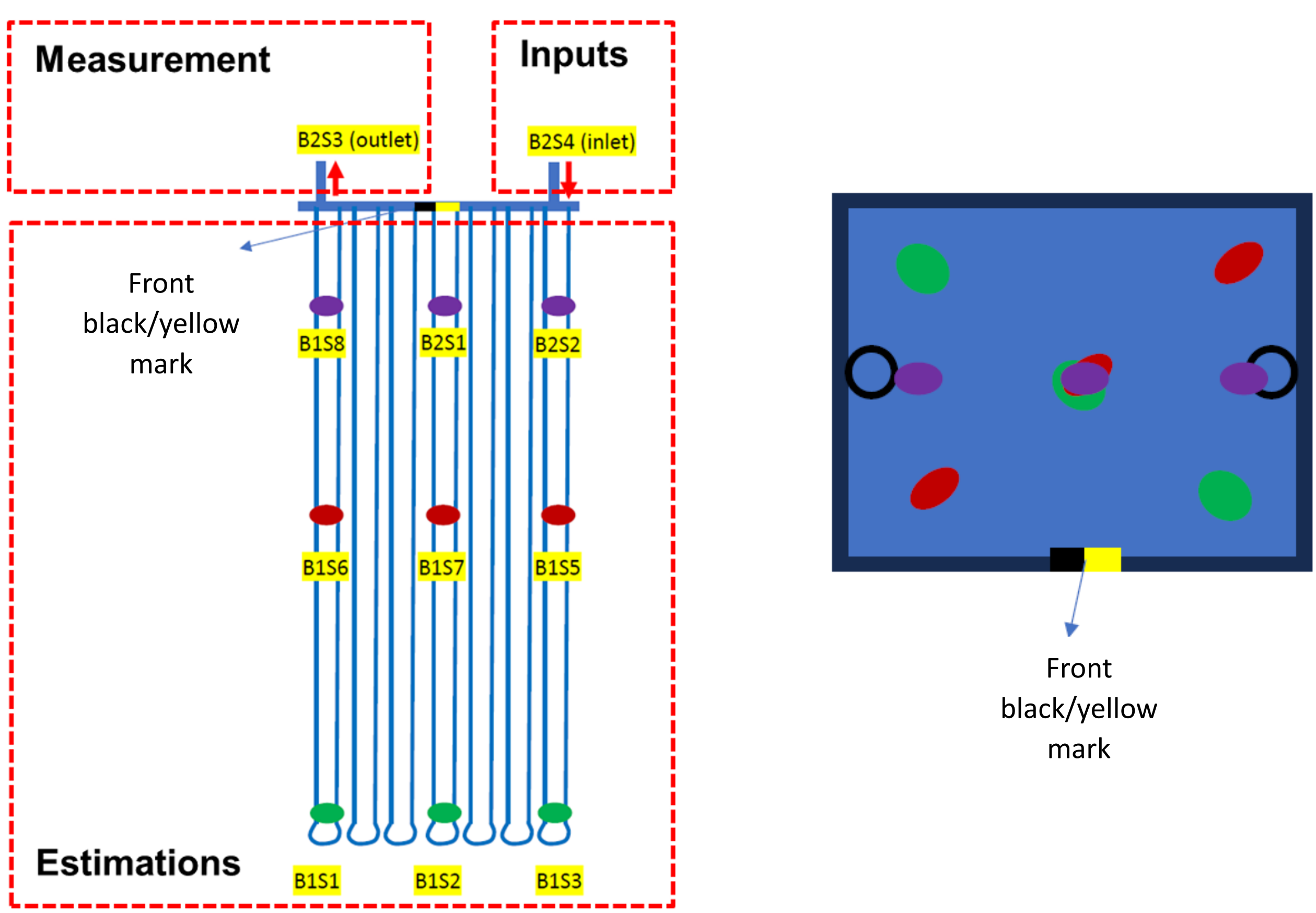

Figure 5: Sensor placement in the PCM HEX during measurement datasets generation, where BxSx are the sensor tag names: (left) side view, (right) top view.

## 3. Results and Discussion

In this section, we present the main simulation results used to validate the eDMDc model for the PCM TES, along with the estimation of PCM temperatures using a KOOPMAN-based observer. For the latter, an observability analysis of the eDMDc model is first conducted, considering the sole available measurement, namely the HTF outlet temperature. Subsequently, the performance of the eDMDc model is evaluated by assessing its ability to predict all internal PCM states. Finally, the capability of the KOOPMAN-based observer to estimate these temperatures using only the available measurements is analyzed.

### 3.1. Observability analysis:

Examination of observability on Eq. (17) gave a rank of 8/10, which means two internal PCM temperature directions are structurally unobservable from the LTES outlet temperature alone. This is expected physical behavior in similar compact PCM-based TESs because:

- Heat diffuses radially.
- Phase front moves spatially.
- Only boundary heat flux is measured (via HTF outlet).

This leads to state indistinguishability: distinct internal temperature fields may yield the same outlet temperature, reflecting a classical inverse heat conduction problem that is inherently ill-posed when only limited boundary measurements are available. Consequently, the proposed KOOPMAN-based observer is expected to capture the overall energy content and track the average PCM temperature, but it cannot uniquely reconstruct the detailed spatial evolution of the phase front. This limitation is not critical, as the primary objective is to provide a reasonably accurate SoC estimation using a minimal set of temperature measurements, rather than to achieve precise reconstruction of all internal temperature states - an objective that is rarely feasible in practical industrial settings.

### 3.2. KOOPMAN (eDMDc) open-loop simulation

In this subsection, we evaluate the eDMDc prediction for the measurement and all states. The simulation results for two charging and discharging cycles following an HTF flow rate of 2 l/min (see Figure 6). The results show that the eDMDc model accurately captures the dominant system dynamics, as evidenced by the close agreement between measured and predicted outlet temperatures over both charging and discharging cycles. The model successfully reproduces the overall trends, switching behavior, and energy exchange driven by the inlet temperature profile. Minor discrepancies appear mainly during sharp transients, where peak smoothing and slight phase shifts are observed, which is typical for reduced-order KOOPMAN-based representations of nonlinear thermal systems.

At the internal state level, the model consistently tracks the temporal evolution of PCM temperatures across all sensors, indicating that the global thermal behavior and energy distribution are well represented. However, deviations in amplitude and local dynamics highlight the intrinsic limitation of reconstructing detailed spatial temperature fields from a single boundary measurement. This confirms the ill-posed nature of the inverse heat conduction problem: while the model is suitable for estimating global quantities such as average temperature or state of charge, it cannot uniquely resolve fine spatial features such as phase front propagation.

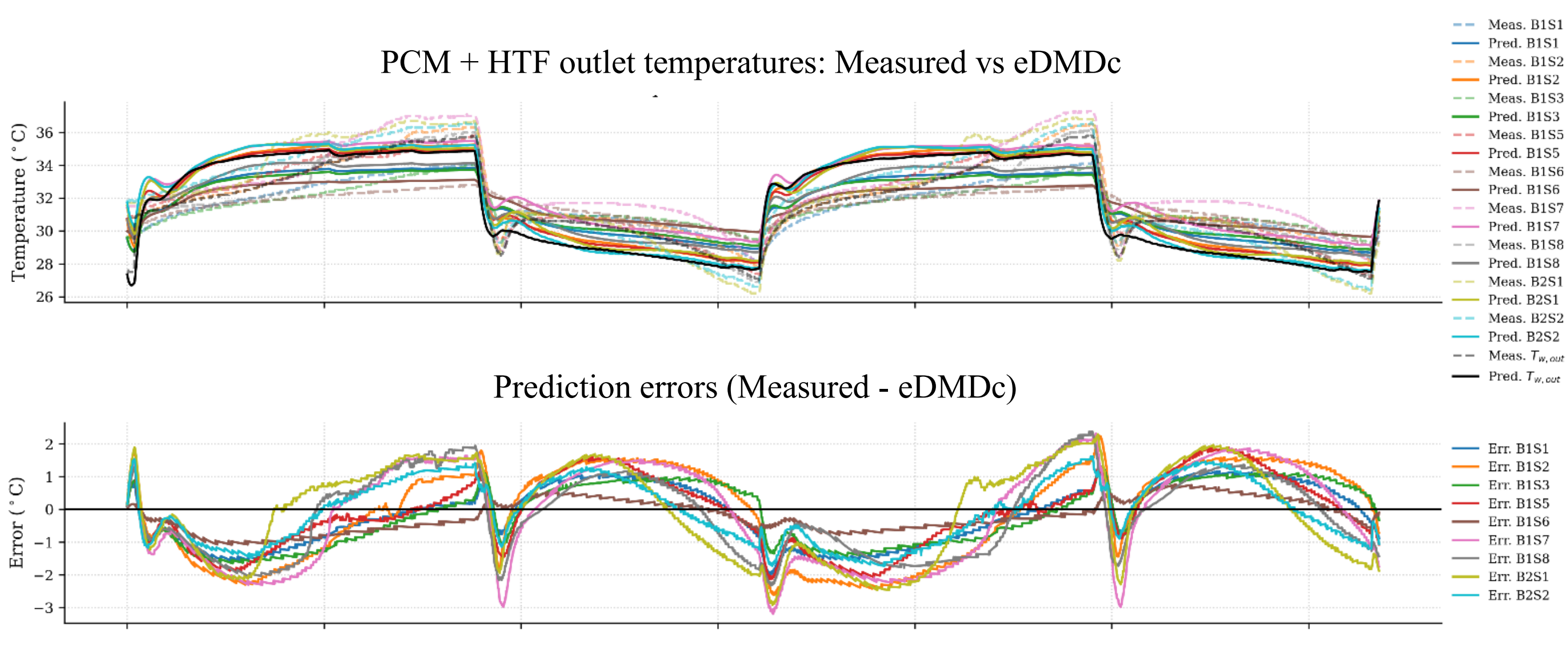

HTF inlet temperature

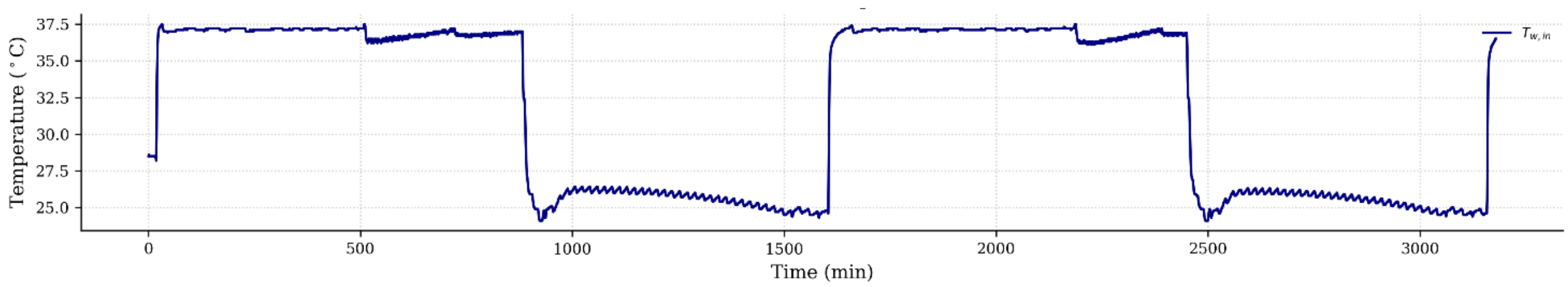

Figure 6: eDMDc prediction performance over two LTES charging and discharging cycles

### 3.3. KOOPMAN-LSO performance analysis

Since the purpose of this work is not the application of eDMDc as a prediction model, as shown in Section 3.2 analysis, rather it is about employing the state observation theory on KOOPMAN to:

- Reconstruct unmeasured states (internal PCM temperatures) using very limited measurements (one in this study).
- Act as a model-based filter to reduce sensor noise, smooth high-frequency fluctuations, and mitigate measurement disturbances.

In this subsection, we will evaluate KOOPMAN-LSO within different operation conditions, namely the HTF flow rate. The assessment will be based on the accuracy of estimating internal PCM temperatures compared with the analysis in Section 3.2. The simulation results are displayed in Figures 7 to 12.

- For HTF flow rate of 2 l/min:

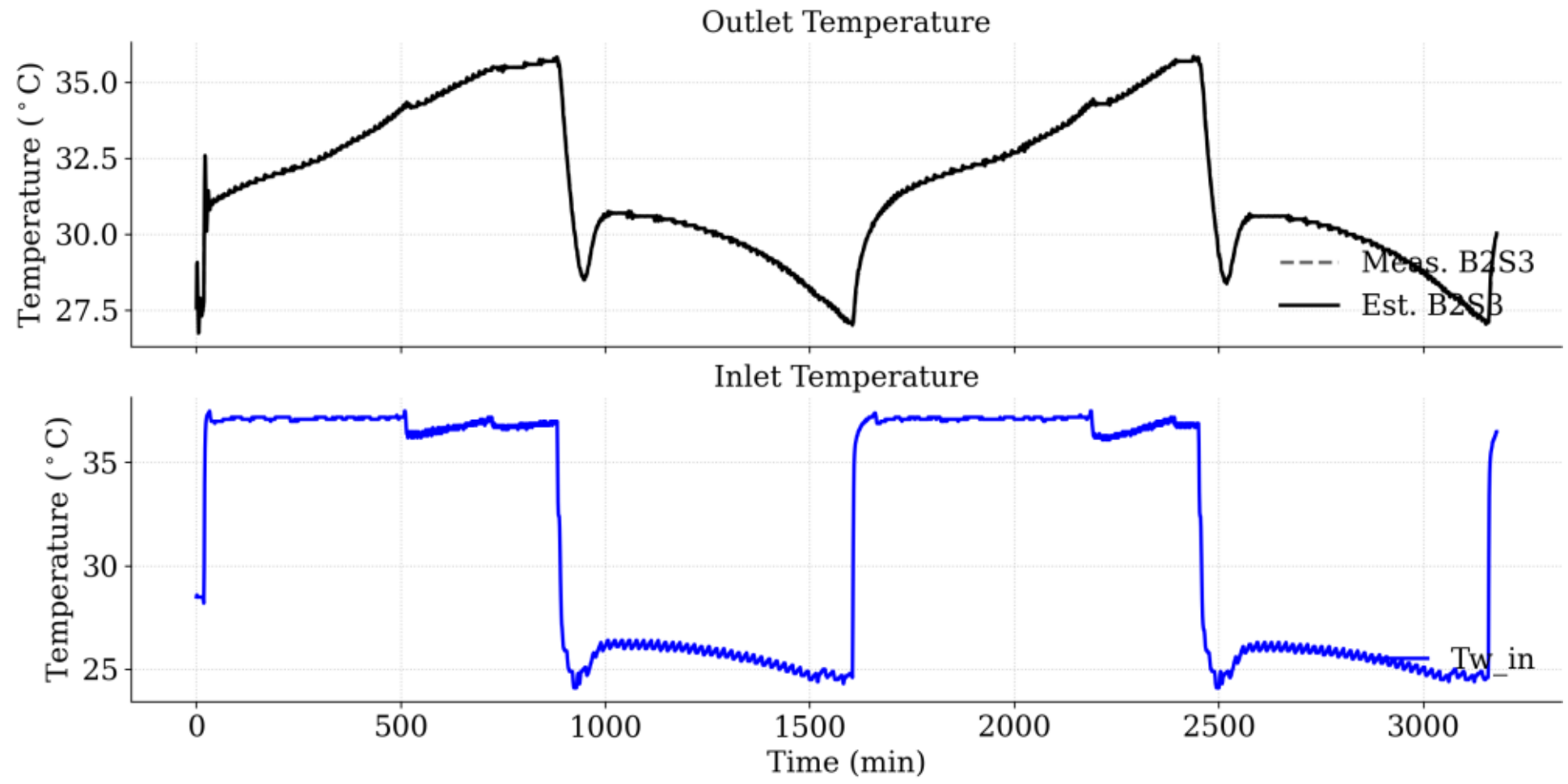

Figure 7: LTES outlet temperature prediction using KOOPMAN-LSO for an HTF flow rate of 2 l/min

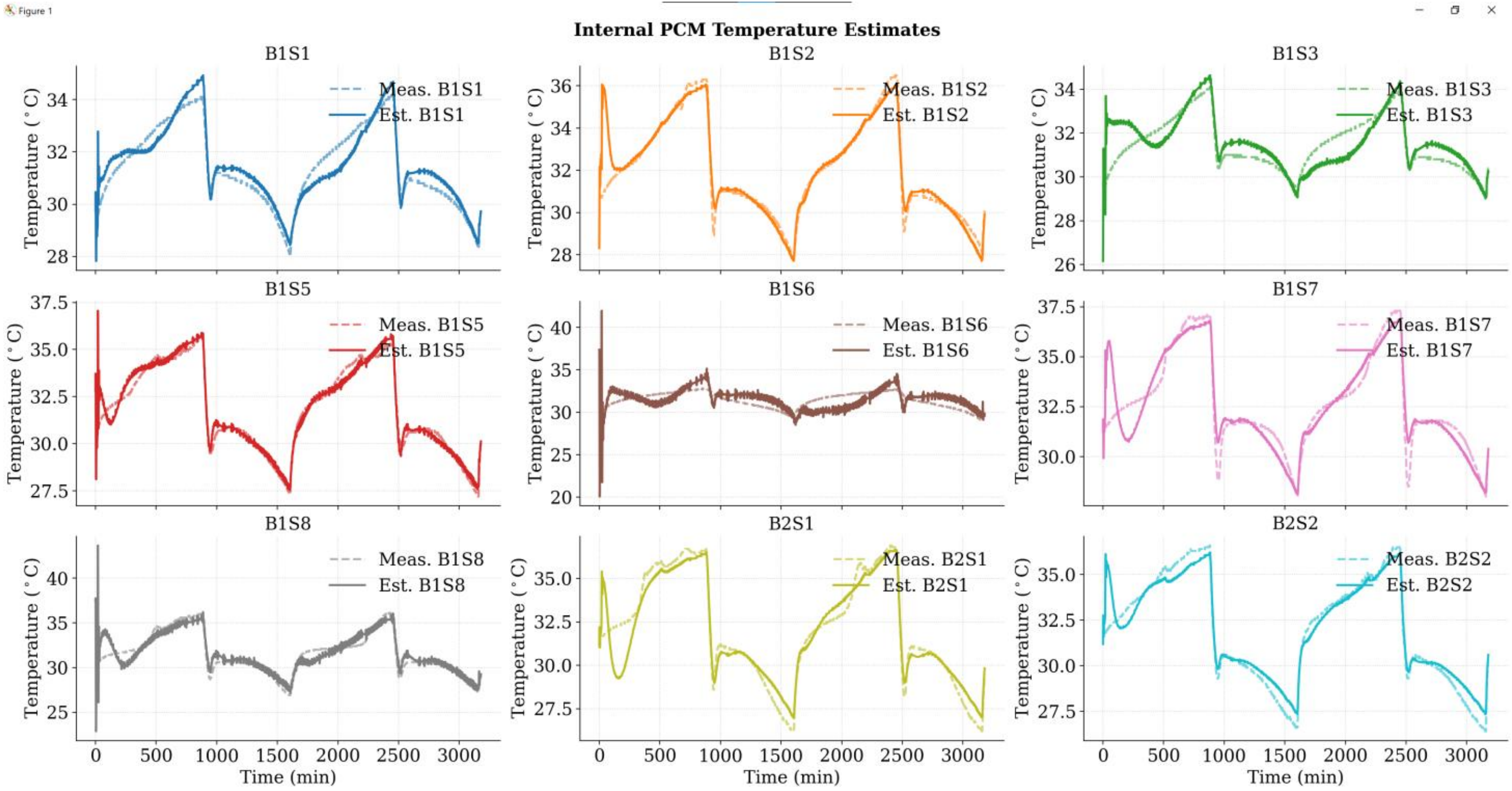

Figure 8: PCM internal temperature estimates using KOOPMAN-LSO for an HTF flow rate of 2 l/min

The LTES outlet temperature plot shows the model's ability to track dynamic temperature changes under a stepwise HTF inlet profile. The KOOPMAN-LSO predictions (solid lines) closely follow the measured temperatures (dashed lines), capturing both the rising and falling trends during charging and discharging cycles. Minor deviations occur during sharp transients, particularly at the start and end of charging phases, likely due to nonlinear effects not fully captured by the linear KOOPMAN operator approximation. Overall, the model maintains accurate predictions within ±1 °C for the majority of the cycle, indicating strong performance for system-level outlet temperature estimation at a constant HTF flow rate of 2 l/min.

The internal PCM temperature estimates across multiple storage nodes (B1S1–B2S2) show that KOOPMAN-LSO effectively captures spatial temperature distributions within the module. Each subplot demonstrates good alignment between measured (dashed) and estimated (solid) temperatures, reproducing both the charge/discharge peaks and the slow cooling phases. Slight discrepancies are observed in nodes with rapid temperature variations (e.g., B1S1, B1S7), reflecting local nonlinearities or thermal lag effects. Overall, the method reliably predicts the internal PCM dynamics, with consistent accuracy across different nodes, supporting its suitability for real-time control and state estimation of latent thermal energy storage at the specified HTF flow rate.

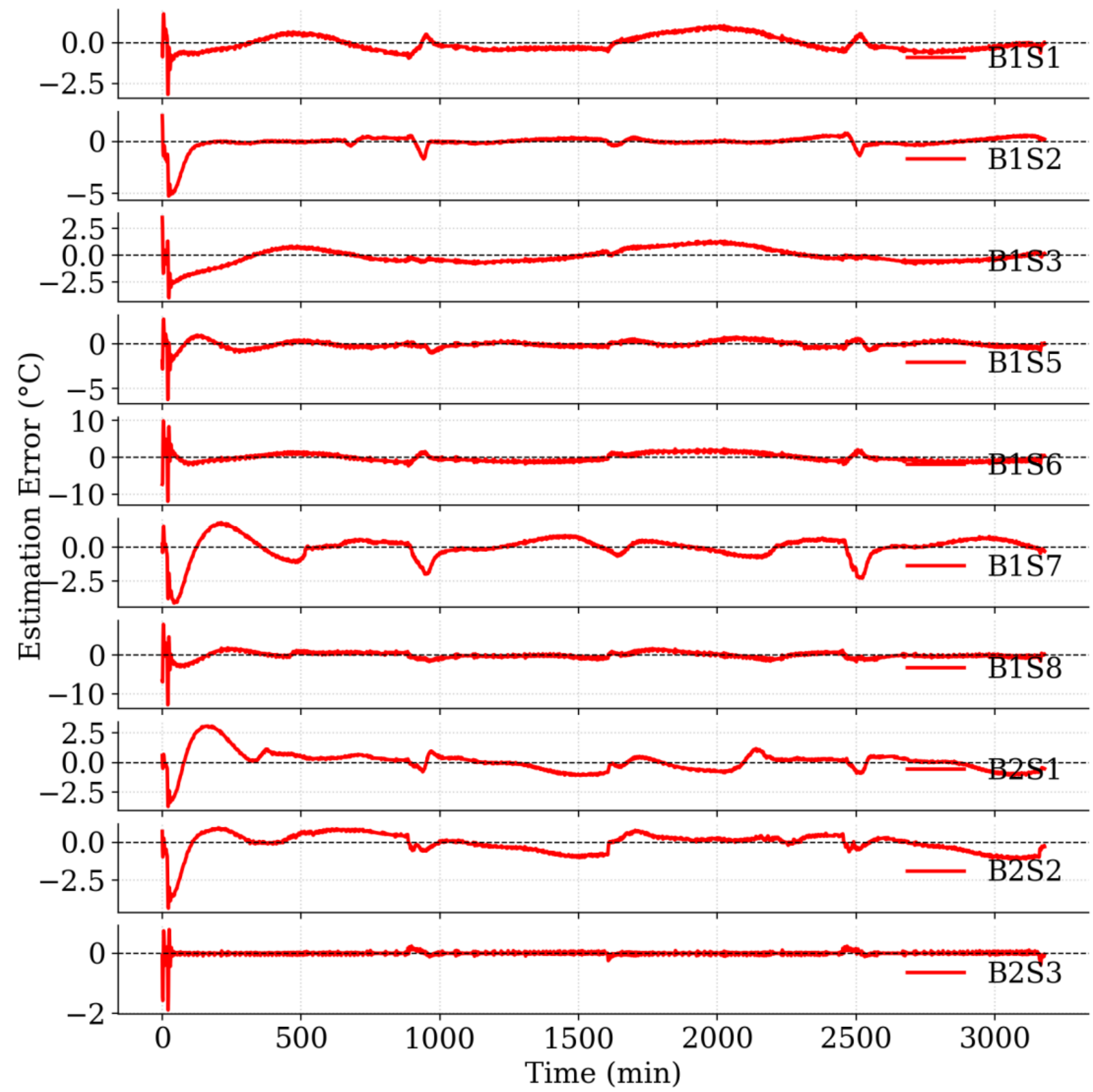

Figure 9: KOOPMAN-LSO estimation errors for an HTF flow rate of 2 l/min

Figure 9 illustrates the dynamics of the estimation error across two charging-discharging cycles. While B2S3 exhibits high estimation accuracy due to direct measurement, the remaining states display acceptable to good estimation performance, with error fluctuations occurring whenever the HTF inlet temperature changes from heating to cooling and vice versa. Some sensors, however, experience larger deviations during transitions between charging and discharging. These discrepancies may stem from the lack of observability of certain states, as discussed in Section 3.1, or from insufficient dependency on the HTF outlet temperature and system inputs to accurately reconstruct the states using the Luenberger observer.

- For HTF flow rate of 4 l/min:

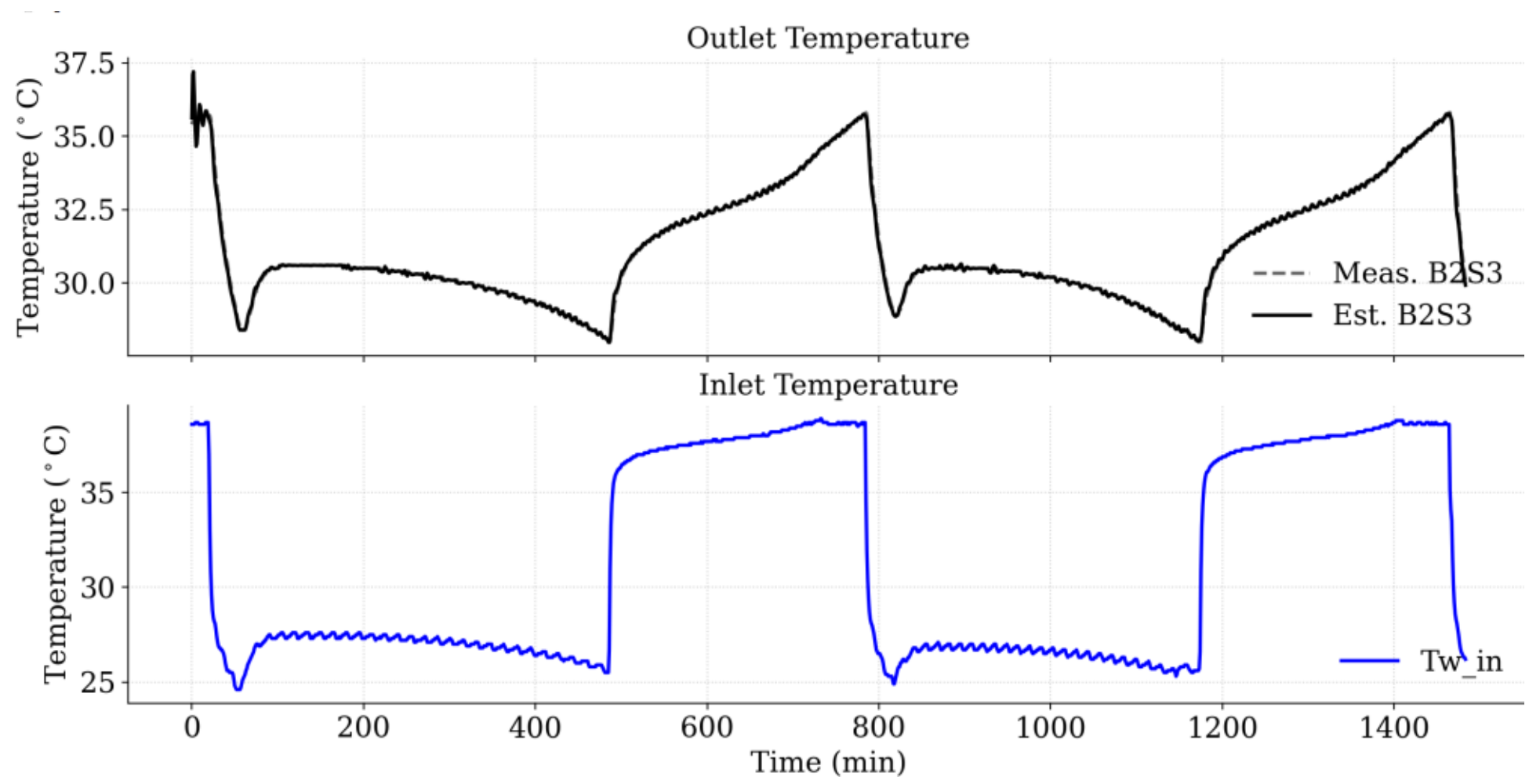

Figure 10: LTES outlet temperature prediction using KOOPMAN-LSO for an HTF flow rate of 4 l/min

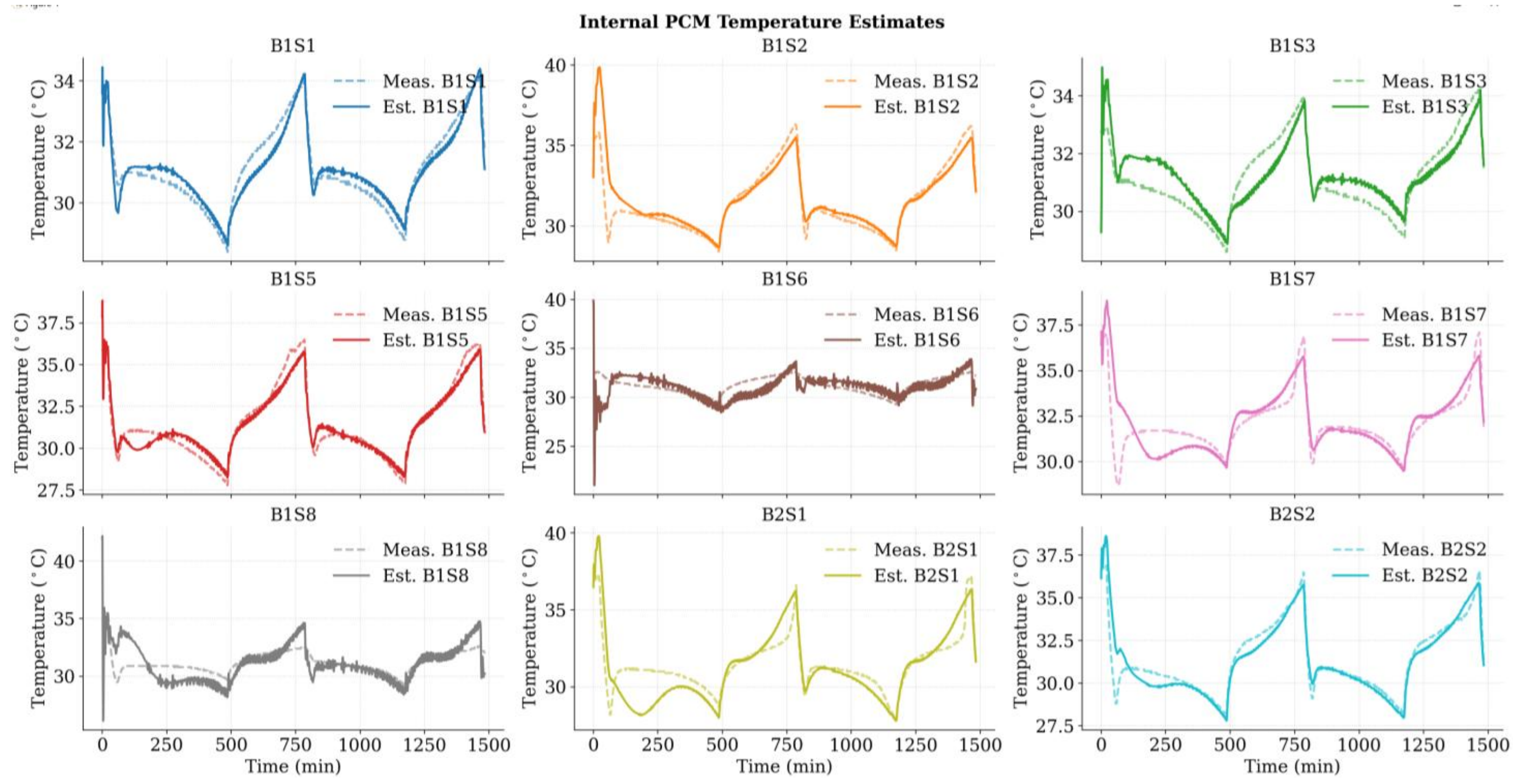

Figure 11: PCM internal temperature estimates using KOOPMAN-LSO for an HTF flow rate of 4 l/min

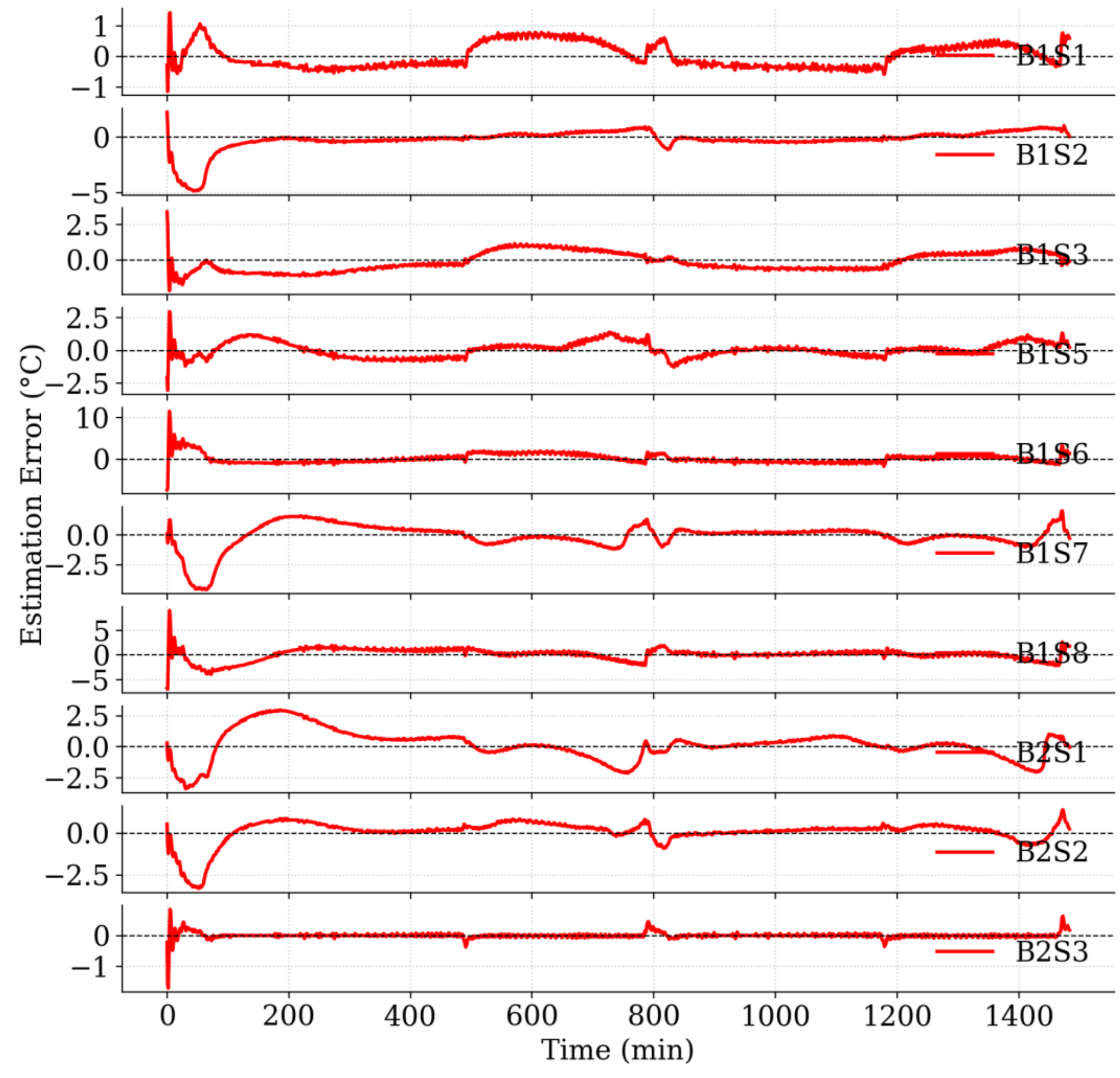

Figure 12: KOOPMAN-LSO estimation errors for an HTF flow rate of 4 l/min

- For HTD flow rate of 8 l/min:

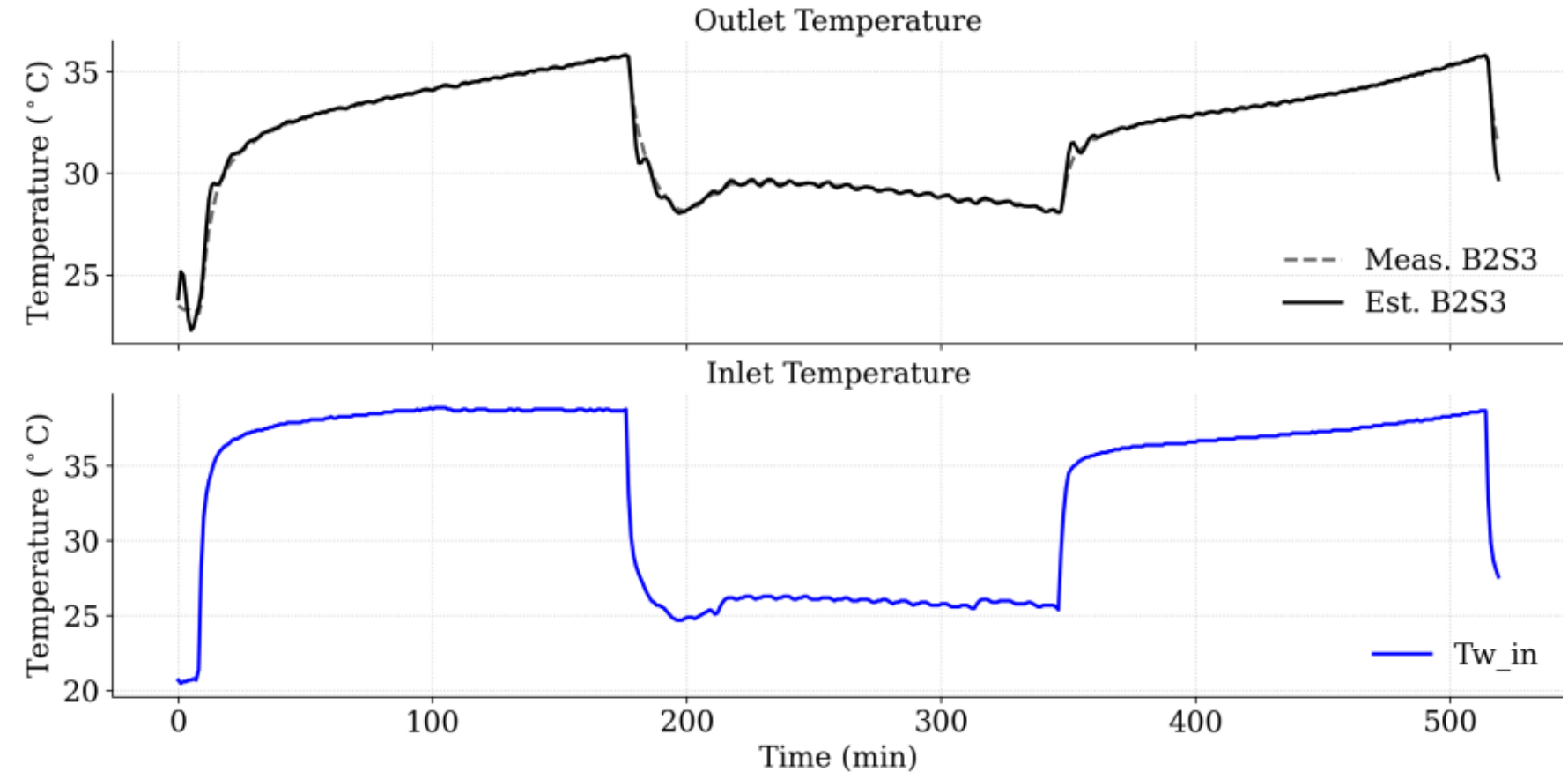

Figure 13: LTES outlet temperature prediction using KOOPMAN-LSO for an HTF flow rate of 8 l/min

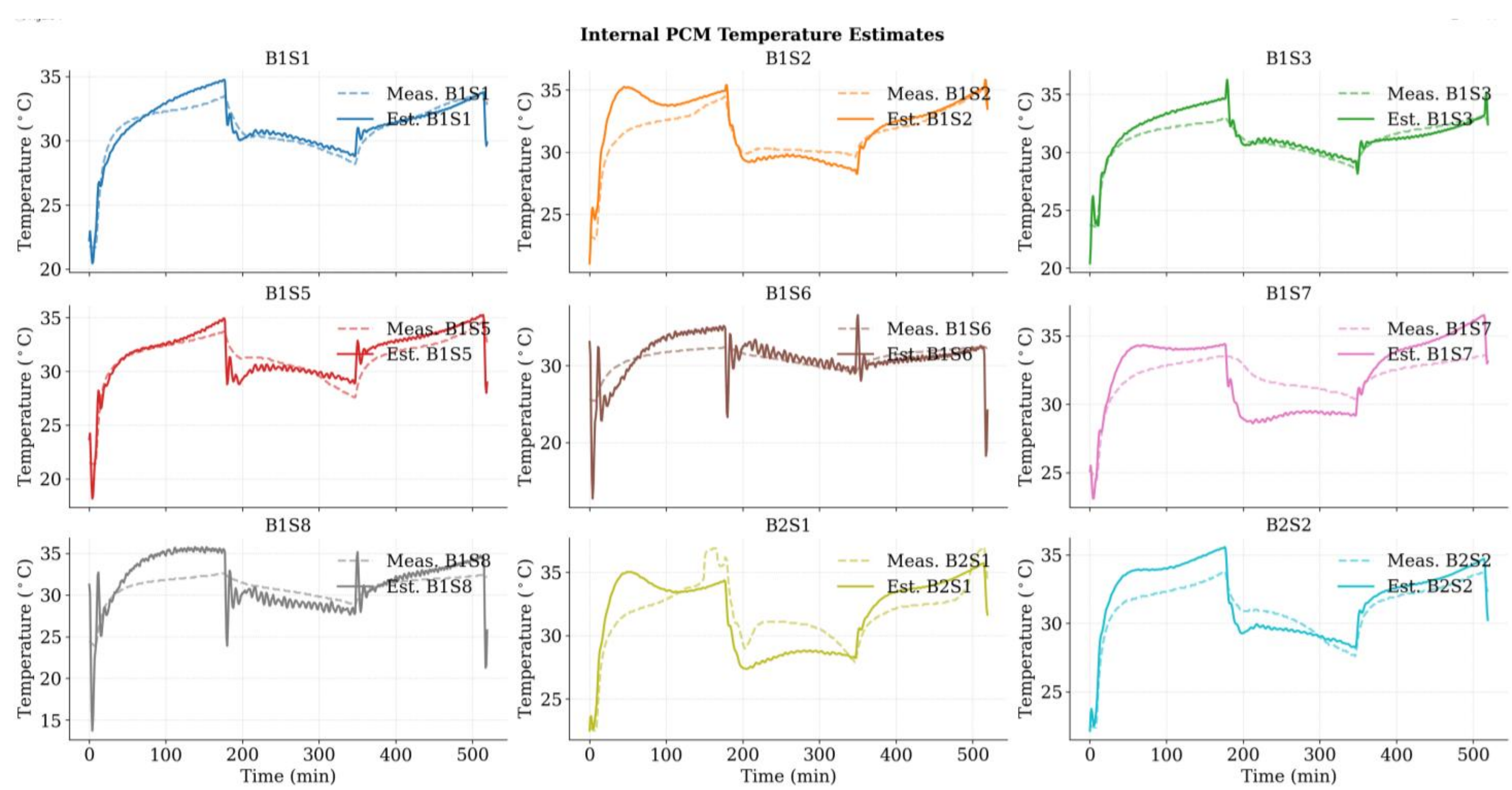

Figure 14: PCM internal temperature estimates using KOOPMAN-LSO for an HTF flow rate of 8 l/min

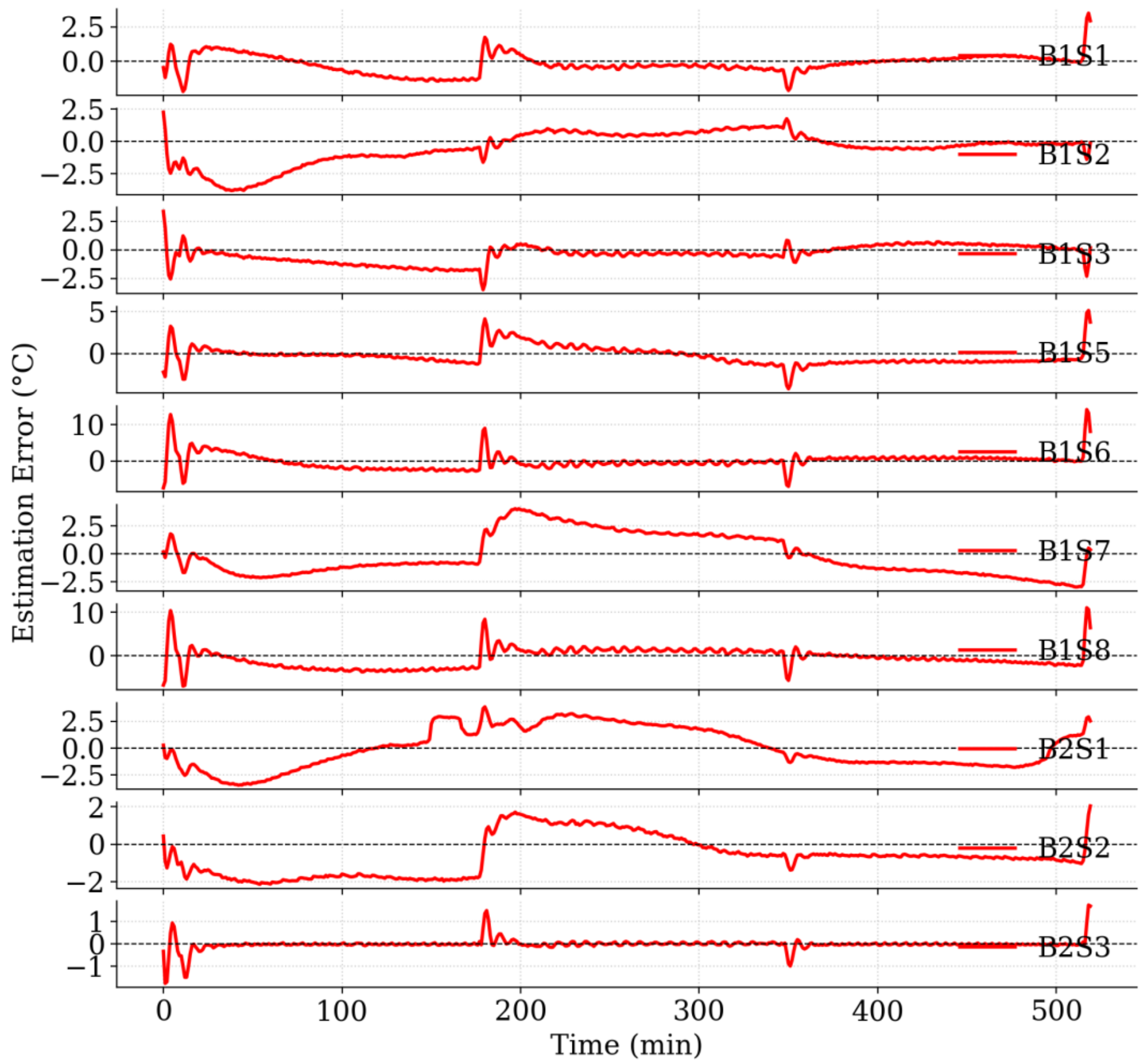

Figure 15: KOOPMAN-LSO estimation errors for an HTF flow rate of 8 l/min

The simulation results presented in Figures 10 through 15 illustrate the performance of the proposed KOOPMAN-LSO framework, particularly highlighting the challenges associated with maintaining estimation accuracy during rapid variations in system inputs. A primary observation from Figures 10 and 13 is the emergence of significant transient periods whenever the HTF inlet temperature undergoes sudden changes - specifically during transitions between charging and discharging cycles. The LSO requires a finite number of iterations to adapt to these step changes, stabilize the estimation error, and force the residual dynamics toward zero. Consequently, temporary estimation deviations are observed at each junction where the HTF temperature fluctuates. This phenomenon is further exacerbated by the system's operating conditions. As depicted in Figure 15, the magnitude of these deviations becomes more pronounced at higher HTF flow rates. The increased flow rate likely intensifies the thermal gradients and reduces the residence time, thereby shortening the effective window for the observer to converge within each phase of the cycle. Table 1 provides a summary of the KOOPMAN-LSO estimation performance across all PCM temperature states under varying operational conditions.

Table 1: KOOPMAN-LSO estimation results for all states under different operation scenarios

| **KOOPMAN-LSO** | | **Sensor tag** | **Average estimates error (RSME) per state - °C** | | |
|---|---|---|---|---|---|
| | | | For 2 l/min flow rate | For 4 l/min flow rate | For 8 l/min flow rate |
| **Estimates** | **States** (PCM temperatures) | B1S1 | 0.4784 | 0.3952 | 0.7125 |
| | | B1S2 | 0.6529 | 0.9568 | 1.3098 |
| | | B1S3 | 0.7231 | 0.6784 | 0.8518 |
| | | B1S5 | 0.4236 | 0.5693 | 1.1039 |
| | | B1S6 | 1.1683 | 1.1755 | 2.1780 |
| | | B1S7 | 0.8145 | 1.0380 | 1.8878 |
| | | B1S8 | 0.9021 | 1.1882 | 2.3072 |
| | | B2S1 | 0.8383 | 1.1955 | 1.8878 |
| | | B2S2 | 0.6592 | 0.6837 | 1.2233 |
| | **Measurement** (HTF outlet temperature) | B2S3 | 0.0819 | 0.1052 | 0.2500 |

### 3.4. Results discussion and limitations

The LSO demonstrates non-uniform performance across the various system states (sensors). While several states are reconstructed with high fidelity, others exhibit persistent tracking errors. This variance in performance can be attributed to two main factors:

1. **Structural unobservability:** As established by the observability analysis in Section 3.1, two specific sensors are fundamentally unobservable within the current system configuration. Without direct or indirect mathematical links to the output measurements, the observer cannot accurately reconstruct these hidden states.
2. **Weak coupling and limited measurement density:** For the remaining sensors that show sub-optimal performance, the issue stems from a weak functional dependency between the limited available measurement points and the specific input variables. In regions of PCM where the thermal coupling to the HTF inlet is physically distant or shielded, the input-output data provides insufficient "excitation" for the KOOPMAN operator to maintain a perfectly accurate state estimation.

These results suggest that while the KOOPMAN-LSO is robust under steady-state or slowly varying conditions, its reliability in highly dynamic LTES applications is constrained by the inherent observability limits and the settling time required following high-frequency input switching.

Despite the mathematical maturity of these methodologies and their initial promise, developing classical observation theory based on data-driven models - such as DMDc and its extended version eDMDc - presents significant challenges for highly nonlinear systems. In such cases, these models often remain valid only within narrow operational windows, thereby restricting their generalizability across diverse dynamic systems and varying operating conditions.

In contrast, the Extended Kalman Filter (EKF) applied to nonlinear models offers a robust alternative to the Luenberger observer on linear data-driven systems, as it is better equipped to capture complex nonlinear patterns across broad operating ranges. To address the difficulty of deriving rigorous first-principles (white-box) mathematical models for PCM-TES systems, Sparse Identification of Nonlinear Dynamics (SINDy) emerges as a powerful data-driven strategy for identifying accurate, interpretable nonlinear representations, which represent a future work to be investigated. This approach was previously validated on an LTES system by Habib et al. [42]; however, its performance has not yet been assessed across a broad range of operating conditions, particularly under varying charging and discharging regimes.

Ultimately, it is imperative to clarify that the primary objective of this framework is not to achieve absolute estimation accuracy across all unmeasured states, particularly those identified as structurally unobservable. Rather, the goal is to establish a streamlined, data-driven methodology that enables the seamless implementation of linear observation theory. By transitioning from a complex nonlinear domain to a KOOPMAN-based linear representation, one can leverage the robust and mature theoretical foundations of linear observers for system stabilization and control. This stands in contrast to nonlinear alternatives, such as EKF, which, despite their accuracy, often lack the global stability guarantees and analytical simplicity inherent to the rich heritage of linear systems engineering.

## 4. Conclusions

This study successfully established a KOOPMAN-based Linear State Observer (KOOPMAN-LSO) framework, demonstrating that physics-informed lifting functions can effectively bridge the gap between nonlinear thermal dynamics and linear estimation theory. By transitioning the system into a higher-dimensional observable space, we enabled the use of a discrete-time Luenberger observer and LQR-based gain tuning - tools that offer far greater analytical simplicity and stability guarantees than traditional nonlinear estimators like the Extended Kalman Filter (EKF).

The application of this framework to a latent thermal energy storage (TES) system yielded robust performance across a range of operational intensities. Quantitative validation showed that the observer can reconstruct the HTF outlet temperature with high precision (RMSE as low as 0.0819 °C) and maintain internal PCM temperature estimates generally below 1.0 °C under standard flow conditions. While estimation errors naturally scaled with increased HTF flow rates - reaching a peak RMSE of 2.3072 °C for specific sensors at 8 l/min - the observer remained stable and convergent throughout the high-frequency switching between charging and discharging cycles.

The performance variances observed across different sensors highlight a critical takeaway: the reliability of data-driven observers is fundamentally bounded by structural observability and the strength of functional dependencies between sparse measurements. Rather than seeking absolute estimation perfection for every unmeasured state, this work provides a computationally efficient methodology for implementing linear control strategies in highly nonlinear environments.

Ultimately, the KOOPMAN-LSO represents a practical middle ground between complex white-box modeling and purely "black-box" data-driven approaches. Future work will investigate the integration of Sparse Identification of Nonlinear Dynamics (SINDy) to further enhance the interpretability of these models and extend their validity across even wider operational windows, paving the way for more advanced, real-time thermal management systems.

**Acknowledgments**

This work was funded by the European Commission under Horizon Europe (Grant 101096789).

**References**

[1] Zalba, B., Marín, J. M., Cabeza, L. F., Mehling, H., "Review on thermal energy storage with phase change: materials, heat transfer analysis and applications," *Applied Thermal Engineering*, 2003. DOI: 10.1016/S1359-4311(02)00192-8

[2] Mehling, H., Cabeza, L. F., *Heat and Cold Storage with PCM*, Springer, 2008. DOI: 10.1007/978-3-540-68557-9

[3] Cabeza, L. F., Castell, A., Barreneche, C., de Gracia, A., Fernández, A. I., "Materials used as PCM in thermal energy storage in buildings," *Renewable and Sustainable Energy Reviews*, 2011. DOI: 10.1016/j.rser.2010.11.018

[4] Voller, V. R., Prakash, C., "A fixed grid numerical modelling methodology for convection–diffusion mushy region phase-change problems," *International Journal of Heat and Mass Transfer*, 1987. DOI: 10.1016/0017-9310(87)90317-6

[5] Shatikian, V., Ziskind, G., Letan, R., "Numerical investigation of a PCM-based heat sink," *International Journal of Heat and Mass Transfer*, 2005. DOI: 10.1016/j.ijheatmasstransfer.2004.10.016

[6] Khodadadi, J. M., Fan, L., "Thermal analysis of PCM heat exchangers," *Journal of Heat Transfer*, 2009. DOI: 10.1115/1.3156814

[7] Dutil, Y., Rousse, D. R., Salah, N. B., Lassue, S., Zalewski, L., "A review on phase-change materials: Mathematical modeling and simulations," *Renewable and Sustainable Energy Reviews*, 2011. DOI: 10.1016/j.rser.2010.06.019

[8] Farid, M. M., Khudhair, A. M., Razack, S. A. K., Al-Hallaj, S., "A review on phase change energy storage," *Energy Conversion and Management*, 2004. DOI: 10.1016/S0196-8904(03)00131-6

[9] De Gracia, A., Cabeza, L. F., "Phase change materials and thermal energy storage for buildings," *Energy and Buildings*, 2015. DOI: 10.1016/j.enbuild.2015.01.002

[10] Barz, T., Seliger, D., Marx, K., Sommer, A., "State and state-of-charge estimation for a latent heat storage," *Control Engineering Practice*, 2018. DOI: 10.1016/j.conengprac.2017.11.007

[11] Bastida, H., de la Cruz-Loredo, I., Ugalde-Loo, C. E., "Nonlinear state observers for latent heat thermal energy storage systems," *Applied Energy*, 2023. DOI: 10.1016/j.apenergy.2022.119900

[12] Shanks, M., Jain, N., "State-dependent Riccati observers for latent thermal energy storage," *ASME Journal of Dynamic Systems, Measurement, and Control*, 2023. DOI: 10.1115/1.4055407

[13] Powell, K. M., Edgar, T. F., "Modeling and control of a solar thermal power plant with thermal energy storage," *Chemical Engineering Science*, 2012. DOI: 10.1016/j.ces.2012.06.012

[14] Wang, Z., Zhang, Y., "State-of-charge estimation of thermal energy storage using data-driven methods," *Energy*, 2018. DOI: 10.1016/j.energy.2018.01.032

[15] Zhang, M., et al., "Data-driven state estimation of latent thermal storage systems," *Energy and AI*, 2021. DOI: 10.1016/j.egyai.2021.100056

[16] Li, G., Hwang, Y., Radermacher, R., "Application of neural networks for thermal energy storage modeling," *Applied Thermal Engineering*, 2012. DOI: 10.1016/j.applthermaleng.2011.10.047

[17] Kong, W., Dong, Z. Y., Jia, Y., Hill, D. J., Xu, Y., Zhang, Y., "Short-term residential load forecasting based on LSTM," *IEEE Transactions on Smart Grid*, 2019. DOI: 10.1109/TSG.2017.2753802 *(methodology adapted in TES estimation literature)*

[18] Korda, M., Mezić, I., "Linear predictors for nonlinear dynamical systems," *Automatica*, 2018. DOI: 10.1016/j.automatica.2018.06.023

[19] Proctor, J. L., Brunton, S. L., Kutz, J. N., "Dynamic mode decomposition with control," *SIAM Journal on Applied Dynamical Systems*, 2016. DOI: 10.1137/15M1013857

[20] Rowley, C. W., Mezić, I., Bagheri, S., Schlatter, P., Henningson, D. S., "Spectral analysis of nonlinear flows," Journal of Fluid Mechanics, 2009. DOI: 10.1017/S0022112009992059

[21] KOOPMAN, B. O. (1931). Hamiltonian systems and transformation in Hilbert space. Proceedings of the National Academy of Sciences, 17(5), 315-318. DOI: 10.1073/pnas.17.5.315

[22] Mezić, I. (2005). Spectral properties of dynamical systems, model reduction and decompositions. Nonlinear Dynamics, 41(1-3), 309-325. DOI: 10.1007/s11071-005-2824-x

[23] Rowley, C. W., Mezić, I., Bagheri, S., Schlatter, P., & Henningson, D. S. (2009). Spectral analysis of nonlinear flows. Journal of Fluid Mechanics, 641, 115-127. DOI: 10.1017/S0022112009992059

[24] Williams, M. O., Kevrekidis, I. G., & Rowley, C. W. (2015). A data-driven approximation of the KOOPMAN operator: Extending dynamic mode decomposition. Journal of Nonlinear Science, 25(6), 1307-1346. DOI: 10.1007/s00332-015-9258-5

[25] Korda, M., & Mezić, I. (2018). Linear predictors for nonlinear dynamical systems: KOOPMAN operator meets model predictive control. Automatica, 93, 149-160. DOI: 10.1016/j.automatica.2018.03.046

[26] Brunton, S. L., Proctor, J. L., & Kutz, J. N. (2016). Discovering governing equations from data by sparse identification of nonlinear dynamical systems. Proceedings of the National Academy of Sciences, 113(15), 3932-3937. DOI: 10.1073/pnas.1517384113

[27] Surana, A. (2016). KOOPMAN operator based observer synthesis for control-affine nonlinear systems. 2016 IEEE 55th Conference on Decision and Control (CDC), 6492-6499. DOI: 10.1109/CDC.2016.7799272

[28] Mamakoukas, G., Castano, M. L., Tan, X., & Murphey, T. D. (2019). Local KOOPMAN operators for data-driven control of robotic systems. Robotics: Science and Systems XV. DOI: 10.15607/RSS.2019.XV.025

[29] Lusch, B., Kutz, J. N., & Brunton, S. L. (2018). Deep learning for universal linear embeddings of nonlinear dynamics. Nature Communications, 9(1), 4950. DOI: 10.1038/s41467-018-07210-0

[30] Li, Z., Kovachki, N., Azizzadenesheli, K., Liu, B., Stuart, A., Bhattacharya, K., & Anandkumar, A. (2020). Fourier neural operator for parametric partial differential equations. International Conference on Learning Representations. DOI: 10.48550/arXiv.2010.08895

[31] Goyal, A., & Panwar, V. (2020). Reduced-order modeling of phase change material-based thermal energy storage systems using proper orthogonal decomposition. Applied Thermal Engineering, 175, 115361. DOI: 10.1016/j.applthermaleng.2020.115361

[32] Zhang, Y., Du, K., He, J., Yang, L., & Li, Y. (2019). A moving boundary model for simulating the thermal behavior of latent heat thermal energy storage. International Journal of Heat and Mass Transfer, 134, 1172-1183. DOI: 10.1016/j.ijheatmasstransfer.2019.01.097

[33] Besançon, G., Voda, A., & Jouffroy, G. (2018). Observer design for nonlinear heat conduction-like phenomena. International Journal of Heat and Mass Transfer, 116, 1237-1245. DOI: 10.1016/j.ijheatmasstransfer.2017.09.088

[34] Alamir, M., & Rouchon, P. (2020). Moving horizon estimation: A guide for the practitioner—with an application to battery state-of-charge estimation. Annual Reviews in Control, 50, 352-368. DOI: 10.1016/j.arcontrol.2020.10.002

[35] Zheng, Y., Li, S., Li, N., & Kar, S. (2021). Distributed continuous-time optimization: A control systems perspective. Annual Reviews in Control, 51, 58-76. DOI: 10.1016/j.arcontrol.2020.11.002

[36] Raissi, M., Perdikaris, P., & Karniadakis, G. E. (2019). Physics-informed neural networks: A deep learning framework for solving forward and inverse problems involving nonlinear partial differential equations. Journal of Computational Physics, 378, 686-707. DOI: 10.1016/j.jcp.2018.10.045

[37] Swischuk, R., Mainini, L., Peherstorfer, B., & Willcox, K. (2020). Projection-based model reduction: Formulations for physics-based machine learning. Computers & Fluids, 198, 104458. DOI: 10.1016/j.compfluid.2019.104458

[38] Li, Q., Dietrich, F., Bollt, E. M., & Kevrekidis, I. G. (2022). Extended dynamic mode decomposition with dictionary learning: A data-driven adaptive spectral decomposition of the KOOPMAN operator. Chaos: An Interdisciplinary Journal of Nonlinear Science, 32(3), 033116. DOI: 10.1063/5.0075906

[39] Nair, A. G., & Goza, A. (2021). Data-driven reduced-order modeling of convective heat transfer with KOOPMAN operators. International Journal of Heat and Mass Transfer, 164, 120586. DOI: 10.1016/j.ijheatmasstransfer.2020.120586

[40] Chen, Y., Vlahostergios, Z., Gatti, D., & Donini, A. (2022). Experimental assessment of data-driven reduced-order models for heat transfer applications. Experimental Thermal and Fluid Science, 131, 110519. DOI: 10.1016/j.expthermflusci.2021.110519

[41] Barz, T., Bres, A., & Emhofer, J. (2022). slPCMlib: A Modelica Library for the Prediction of Effective Thermal Material Properties of Solid/Liquid Phase Change Materials (PCM). In Proceedings of Asian Modelica Conference 2022 (pp. 63-74). Linkoping University Electronic Press. DOI: https://doi.org/10.3384/ecp19363.

[42] Mustapha Habib, Youssef Elomari, Felix Hochwallner, Adam Buruzs, Tilman Barz, and Qian Wang. Extended Kalman filter on sparse identification of nonlinear systems: application to the SoC estimation of a phase change material-based energy storage. Energy Conversion and Management: X 27 (2025) 101199